\documentclass[a4paper,fleqn]{cas-sc}   
\usepackage[numbers]{natbib}
\usepackage{amsmath}
\usepackage{graphicx} 
\usepackage{xcolor}
\usepackage{lineno}
\usepackage{units,siunitx}
\usepackage{fancyvrb}
\usepackage{rotating}
\usepackage{threeparttable}
\usepackage{subcaption}
\usepackage{caption}
\usepackage{amsmath}
\usepackage{array} 
\usepackage{multirow}
\usepackage{setspace}
\newcommand{\vectorstyle}[1]{\vec{\mathbf{#1}}}

\begin{document}
\let\WriteBookmarks\relax
\def\floatpagepagefraction{1}
\def\textpagefraction{.001}
\shorttitle{NJOY+NCrystal: an open-source tool for creating thermal neutron scattering libraries}

\title [mode = title]{NJOY+NCrystal: an open-source tool for creating thermal neutron scattering libraries with mixed elastic support}

   
\author[1]{Kemal Rami\'{c}}
\ead{Kemal.Ramic@ess.eu}

\address[1]{European Spallation Source ERIC, Lund, Sweden}

\author[1]{Jose Ignacio {Marquez Damian}}
\ead{marquezj@ess.eu}

\author[1]{Thomas Kittelmann}
\ead{Thomas.Kittelmann@ess.eu}

\author[1]{Douglas D. {Di Julio}}
\ead{Douglas.DiJulio@ess.eu}

\author[2]{Davide Campi}
\ead{davide.campi@epfl.ch}

\address[2]{University of Milano-Bicocca, Milano, Italy}

\author[2]{Marco Bernasconi}
\ead{marco.bernasconi@unimib.it}

\author[2]{Giuseppe Gorini}
\ead{giuseppe.gorini@unimib.it}

\author[1]{Valentina Santoro}
\ead{Valentina.Santoro@ess.eu}

\begin{abstract}
 In this work we present NJOY+NCrystal, a tool to generate thermal neutron scattering libraries with support for coherent and incoherent elastic components for crystalline solid materials. This tool, which is a customized version of NJOY, was created by modifying the nuclear data processing program NJOY to call the thermal scattering software library NCrystal, and includes a proposed change in the ENDF-6 format to store both the coherent and incoherent elastic components. Necessary changes to enable this format in NJOY, as well as to sample it in the OpenMC Monte Carlo code, are detailed here. Examples of materials that are coherent-dominant, incoherent-dominant, and mixed elastic scatterers are presented, as well as the creation of novel libraries for MgH$_2$ and MgD$_2$, that are under consideration as advanced neutron reflectors in the HighNESS project at the European Spallation Source. NJOY+NCrystal greatly simplifies the process to generate thermal scattering libraries (TSL) and this is exemplified with 213 new and updated TSL evaluations.

\end{abstract}


\maketitle

\begin{spacing}{2.0}

\section{Introduction}
The European Spallation Source, which is under construction in Lund, Sweden, will be the most powerful neutron source in the world. The source will utilize a butterfly moderator system, placed above the spallation target, to produce the brightest source of neutrons, which will be used by 15 instruments for a wide range of experiments. 
Recently, a project named HighNESS, \cite{wwwhighness,santoro2020development}, has been initiated in order to investigate the design of a second moderator system, which would utilize a high-intensity neutron source with an emphasis on cold, very cold and ultra-cold neutrons. The project includes both the usage of novel moderator and reflector materials, in addition to enhanced techniques for producing and shaping neutron beams. Part of the effort is the creation of new accurate nuclear data describing the interaction of neutrons with these novel materials.

The design of neutron reactors, neutron sources, and nuclear systems requires calculations of neutron distributions using radiation transport codes. The source for nuclear data used in radiation transport codes, such as \cite{ref_MCNP6,ROMANO201590,doi:10.1080/00223131.2017.1419890}, are the evaluated nuclear data files (ENDF). ENDF files include sub-libraries for different kinds of particles, in particular thermal neutron scattering data in form of thermal scattering laws (TSL) and written in ENDF-6 format \cite{osti_1425114}. In ENDF format, File 7 stores the thermal neutron scattering data in two sections: one for elastic and one for inelastic scattering. In general, the scattering cross section of neutrons can be decomposed into four parts: coherent elastic, incoherent elastic, coherent inelastic, and incoherent inelastic scattering. Elastic scattering represents the scattering of a neutron without the exchange of energy with the target material in the laboratory frame of reference, while inelastic scattering represents the scattering of a neutron where it either gains or loses energy. Coherent scattering contains the interference terms of the scattering and therefore is sensitive to the structure of the material, whereas incoherent elastic scattering depends on the self correlation for each atom at different times. 

At the moment, in the ENDF-6 format, the elastic section stores either the coherent elastic or incoherent elastic cross section only. In the inelastic section, either the incoherent inelastic cross section is stored in the incoherent approximation or the inelastic is stored as a sum of incoherent and coherent inelastic parts calculated from the coherent one-phonon approximation. To overcome the limitation in the elastic section, a change was recently proposed \cite{ref_Zerkle} for the thermal scattering format, which will in this paper be referred to as the ``mixed elastic format''. The new format proposes to store both the coherent and incoherent elastic cross sections in the elastic section, one after another one, in the same format as in which they were stored before individually. 

Thermal scattering files can be produced with the LEAPR module of NJOY \cite{ref_njoy2016}, which is a nuclear data processing code developed by Los Alamos National Laboratory and it is an industry standard in its class of codes. As mentioned, the LEAPR module can be used for creation of TSL files, and although it has been widely used to do so, it has some limitations. Mainly, LEAPR has a limited support for coherent elastic scattering physics through few hard-coded materials (seen in Table \ref{tab:iel_leapr}), the inelastic component can be only calculated in the incoherent approximation, and it is limited by the physics that can be stored in ENDF format. On the other hand, the NCrystal library \cite{CAI2020106851, CAI2019400} has an extensive treatment of the calculation of thermal neutron scattering cross sections directly but cannot generate ENDF-6 formatted files. NCrystal supports a range of physics, including coherent and incoherent elastic scattering as well as the inelastic scattering, in a wide range of materials, including powders, mosaic single crystals, layered single crystals and liquids, and it also contains a data library covering many materials important at neutron scattering facilities. In the current ENDF-6 format, most of the physics enabled in NCrystal could not be stored in the tsl-ENDF file, hence in this work we propose to combine these two tools, using NCrystal to generate the microscopic data that is later used by NJOY to produce thermal scattering libraries for poly-crystalline (or sometimes referred to as crystals treated in the powder approximation) materials. Additionally, we include modifications to support the proposed mixed elastic format for tsl-ENDF files and produce thermal scattering .ACE files, in the here proposed format with both coherent and incoherent elastic scattering components. With NJOY+NCrystal, the user isn't anymore limited to materials with hard-coded options, but any solid poly-crystalline material can be calculated with support for coherent and incoherent elastic components in the current ENDF format, as well as in the proposed mixed elastic format. The Monte Carlo code OpenMC \cite{ROMANO201590} was modified as well to support this new format of .ACE files, which makes this a complete implementation, starting from microscopic calculations and ending with Monte Carlo sampling.


\section{Theory}

As presented in \cite{CAI2020106851}, the scattering of neutrons can be represented in terms of the double differential scattering cross section as:
\begin{equation}\label{eq:DDXS}
    \frac{d^2\sigma_{\vec{k}\Rightarrow\vec{k}'}}{d\Omega'dE'}=\frac{k'}{k}S(\vec{Q},\omega)
\end{equation}
where $\sigma$ is the cross section, d$\Omega'$ is the scattering angle, dE' is the scattered energy of neutron, $k$ and $k'$ are wavenumbers for the incoming and scattered neutron, $\hbar\vec{Q}$ is the momentum transfer, equal to $\hbar\vec{Q}=\hbar\vec{k'}-\hbar\vec{k}$. $S(\vec{Q},\omega)$ is the scattering function, which for a crystalline system is divided into an equivalent subsystem and can be represented as: 
\begin{equation}\label{eq:sqw}
    S(\vec{Q},\omega)=\frac{1}{2\pi\hbar}\sum^{N_{sub}}_{jj'=1}\overline{b_jb_{j'}}\int^{\infty}_{-\infty} \langle j,j' \rangle e^{-i \omega t} dt
\end{equation}
where $N_{sub}$ represents the subsystem size, $\overline{b_jb_{j'}}$ are the average scattering lengths over the subsystem, and $\langle j,j' \rangle$ is the expectation value at thermal equilibrium which correlates the position of the nucleus $j$ at time $t$ with the position of the nucleus $j$' at time 0 and it has form $\langle e^{-i\vec{Q}\cdot \vectorstyle{R}_{j'}(0)}e^{-i\vec{Q}\cdot \vectorstyle{R}_{j}(t)} \rangle$. The scattering function of such a system can then be represented as a sum of coherent and incoherent parts:

\begin{equation}\label{eq:sum_sqw}
    S(\vec{Q},\omega)=S_{coh}(\vec{Q},\omega)+S_{inc}(\vec{Q},\omega)
\end{equation}
\begin{equation}\label{eq:sqw_coh}
    S_{coh}(\vec{Q},\omega)=\frac{1}{2\pi\hbar}\sum^{N_{sub}}_{jj'=1}\overline{b_j}\cdot \overline{b_{j'}}\int^{\infty}_{-\infty} \langle j,j' \rangle e^{-i\omega t} dt
\end{equation}
\begin{equation}\label{eq:sqw_incoh}
    S_{inc}(\vec{Q},\omega)=\frac{1}{2\pi\hbar}\sum^{N_{sub}}_{j=1}\left (\overline{b^2_j} - \left(\overline{b_{j}}\right)^2 \right)\int^{\infty}_{-\infty} \langle j,j \rangle e^{-i\omega t} dt.
\end{equation}
where $\langle j,j' \rangle$ is dependent on positions of the nuclei in a crystal. This correlation can be further simplified by applying the \emph {harmonic approximation} which consists in separating the position operator in displacements $\vectorstyle{u}$ from equilibrium positions $\vec{d}$. The resulting correlation between displacements can be expanded in a Taylor series:

\begin{equation}\label{eq:corrl}
    e^{\langle(\vec{Q}\cdot \vectorstyle{u}_{j'}(0))(\vec{Q}\cdot \vectorstyle{u}_{j}(t)) \rangle} = \sum_{n=0}^{\infty} \dfrac{1}{n!} \left( \langle(\vec{Q}\cdot \vectorstyle{u}_{j'}(0))(\vec{Q}\cdot \vectorstyle{u}_{j}(t)) \rangle \right)^n
\end{equation}

The first term in this expansion corresponds to elastic scattering. The correlation in this term is known as the Debye-Waller function:

\begin{equation}\label{eq:debye_waller_function}
    W_j(\vec{Q})\equiv\tfrac{1}{2}\langle (\vec{Q}\cdot \vectorstyle{u}_j(0))^2 \rangle.
\end{equation}

For isotropic systems (i.e. polycrystalline samples), integration over all orientations results in a Q-dependent Debye-Waller function:
\begin{equation}\label{eq:isotropic_debye_waller_factor}
    2W(Q)=\tfrac{1}{3}Q^2\langle u^2 \rangle,
\end{equation}
where $\langle u^2 \rangle$ is the mean-squared displacement (MSD).

\subsection{Incoherent elastic scattering}
For elastic scattering, with $k=k'$, the incoherent elastic double differential scattering cross section is equal to:
\begin{equation}\label{eq:DDXS_incoherent_elastic}
\begin{split}
    \frac{d^2\sigma_{\vec{k}\Rightarrow\vec{k'}}^{el,inc}}{d\Omega'dE'}&=\frac{k'}{k}S^{inc}_{el}(\vec{Q},\omega)\\
    &=\sum^{N_{sub}}_{j=1}\left (\overline{b^2_j} - \left(\overline{b_{j}}\right)^2 \right)e^{-2W_j(\vec{Q})}\delta(\hbar\omega).
\end{split}
\end{equation}
\noindent where the bound incoherent cross section, $\sigma_b^{inc}$, is equal to $\sigma_{b}^{inc}=4\pi(\overline{b^2_j} - (\overline{b_{j}})^2)$. 
In NJOY, the incoherent elastic cross section is calculated in the THERMR module as:
\begin{equation}\label{eq:thermr_incoh_elas_xs}
    \frac{d^2\sigma_{\vec{k}\Rightarrow\vec{k'}}^{el,inc}}{d\Omega'dE'}=\frac{\sigma_b^{inc}}{4\pi} e^{-2W(Q)}
\end{equation}
with:
\begin{equation}\label{eq:thermr_ncrystal_dw}
    2W(Q) = 2 W_\text{ENDF} E (1-\mu)
\end{equation}
where E is the incident energy of the neutron, $\mu$ is the cosine of the scattering angle, and $W_\text{ENDF}$ is the Debye-Waller integral in eV$^{-1}$ and is computed from the phonon spectrum in the LEAPR module of NJOY:
\begin{equation}\label{eq:thermr_dw_phonon_dos}
    W_\text{ENDF} = \dfrac{\hbar^2}{2 M kT} \int_0^{\infty} \dfrac{\rho(\omega)}{\hbar \omega / kT} \coth \left(\hbar \omega/(2 kT)\right) \mathrm d \omega
\end{equation}

$W_\text{ENDF}$ and $\sigma_b^{inc}$ are stored in the incoherent elastic section of the ENDF file by LEAPR if the incoherent elastic cross section is present. The integrated incoherent elastic cross section is equal to:
\begin{equation}\label{eq:XS_incoh_elas}
    \sigma_{el,inc}(E)=\frac{\sigma_b^{inc}}{2} \left\{\frac{1-e^{-4W_\text{ENDF}E}}{2W_\text{ENDF}E}\right\}.
\end{equation}

\subsection{Coherent elastic scattering}

The scattering function for the coherent elastic component can be written as:
\begin{equation}\label{eq:sqw_coherent_elastic}
    S^{coh}_{el}(\vec{Q},\omega)=\frac{(2\pi)^3\delta(\hbar\omega)}{V_{uc}}\sum_{hkl}\delta(\vec{Q}-\vec{\tau}_{hkl})|F(\vec{\tau}_{hkl})|^2,
\end{equation}
where $V_{uc}$ is the volume of the unit cell,$|F(\vec{\tau}_{hkl})|$ is the form factor of the unit cell, and $\vec{\tau}_{hkl}$ is a point in the reciprocal lattice. In the powder approximation, where the crystal grains appear with uniformly randomised orientations, the integrated coherent elastic scattering cross section can be written as: 
\begin{equation}\label{eq:XS_coherent_elastic}
    \sigma_{el,coh}^{hkl}(E)=\frac{\pi^2 \hbar^2}{m_n E V_{uc}}\sum^{E_{hkl} \leq E}_{hkl}d_{hkl}|F(\vec{\tau}_{hkl})|^2,
\end{equation}
where $E_{hkl} = \dfrac{\hbar^2\tau_{hkl}^2}{8 m_n} $ is the energy threshold of the Bragg edge caused by the family of planes with Miller indices $(hkl)$, and $d_{hkl}$ is the d-spacing of the planes. In the ENDF-6 format this information is stored as pairs $\left(E_{hkl}, E_{hkl} \sigma_\text{el,coh}^{hkl}\right)$.


\subsection{Inelastic scattering}

The $n \geq 1$ terms in the expansion of eq. \ref{eq:corrl} correspond to inelastic scattering. In NJOY+NCrystal this is handled by the LEAPR module, in which the double differential scattering cross section is defined in the incoherent approximation as:
\begin{equation} \label{eq:doubledifferential}
    \frac{d^2\sigma_{\vec{k}\Rightarrow\vec{k'}}^{inelas}}{d\Omega'dE'}=\frac{\sigma_b}{4\pi kT}\sqrt{\frac{E'}{E}} S^\text{LEAPR}(\alpha, \beta),
\end{equation}
where $\mu$ is the cosine of the scattering angle, and $S^\text{LEAPR}(\alpha, \beta)$ is the scattering function. This is different from the previous definition in eq. \ref{eq:DDXS} because in the incoherent approximation the bound atom scattering cross section is factored out. The variables $\alpha$ and $\beta$ are related respectively to the momentum transfer and energy transfer:

\begin{equation} \label{eq:AlphaAndBeta}
	\alpha=\frac{E'+E-2\mu\sqrt{E'E}}{Ak_BT},\quad \beta=\frac{E'-E}{k_BT}.
\end{equation}

where \emph{A} is the ratio of the mass of the scattering atom to the neutron mass. 
In the incoherent and Gaussian approximation the $S(\alpha, \beta)$ is defined as:
\begin{equation} \label{eq:SAlphaAndBeta}
	S^\text{LEAPR}(\alpha,\beta)=\frac{1}{2\pi}\int_{-\infty}^{\infty}e^{i\beta\hat{t}}e^{-\gamma(\hat{t})}d\hat{t},
\end{equation}
where:
\begin{equation} \label{eq:gammat}
	\gamma(\hat{t})=\alpha\int_{-\infty}^{\infty}P(\beta)[1-e^{-i\beta\hat{t}}]e^{-\beta/2}d\beta,
\end{equation}
with:
\begin{equation} \label{eq:PBeta}
	P(\beta)=\frac{\rho(\beta)}{2\beta\sinh(\beta/2)},
\end{equation}
where $\rho(\beta)$ is the phonon spectrum, and $\hat{t}$ is the time measured in units of $\hbar/kT$. 

The total inelastic cross section is obtained by integrating eq. \ref{eq:doubledifferential} in outgoing energy and scattering angle:
\begin{equation}\label{eq:XS_inelastic}
    \sigma_{inel}(E)=\int_{4\pi} \mathrm{d}\Omega' \int_{0}^\infty \mathrm{d}E'\, \frac{d^2\sigma_{\vec{k}\Rightarrow\vec{k'}}^{inelas}}{d\Omega'dE'}
\end{equation}

%
%

\subsection{Total cross section}

The total interaction probability for an incident energy $E$ will be given by the sum of the absorption cross section (representing the probability of all events that are not scattering), the inelastic cross section (eq. \ref{eq:XS_inelastic}, and the coherent (eq. \ref{eq:XS_coherent_elastic} and incoherent (eq. \ref{eq:XS_incoh_elas}) elastic cross sections, if present:

\begin{equation}\label{eq:XS_total}
    \sigma_{tot}(E)=\sigma_{abs}(E)+ \sigma_{inel}(E) + \sigma_{el,coh}(E) + \sigma_{el,inc}(E)
\end{equation}

\section{Implementation}\label{sec_implementation}


NJOY+NCrystal was motivated mainly by a need to implement the mixed elastic format proposal in an open source code, such as NJOY. Additionally, the motivation was to provide an implementation of both a proper and user-friendly handling of the coherent and incoherent elastic cross section calculations, in both the current and proposed format, in NJOY by utilizing NCrystal. Therefore, NJOY+NCrystal is a customized version of NJOY that relies on NCrystal to provide the necessary information for the calculations of the coherent and incoherent elastic components, whereas the inelastic component will be calculated in LEAPR module of NJOY+NCrystal. Additionally, both NJOY and NCrystal utilize CMake and GitHub Repository's for the installation, hence making the integration smoother. 

The changes to NJOY can be split into two parts. The first part is to provide an interface between LEAPR and NCrystal, described briefly in \ref{subsection:leapr_modification}, so that LEAPR can calculate coherent and incoherent elastic cross section components for any crystalline material. The calculated elastic components are then stored in the tsl-ENDF file for both the current and proposed mixed elastic format. The second part consists of implementing changes to the THERMR and ACER modules of NJOY so that the new mixed elastic format can be read and handled properly in order to produce .ACE files used by Monte Carlo programs to sample neutron scattering events. The changes to the OpenMC Monte Carlo transport code are presented in Section \ref{openmc_modifications}, as an example of the sampling here proposed mixed elastic .ACE format.

\subsection{LEAPR modifications}\label{subsection:leapr_modification}

In the current LEAPR module of NJOY, which is used to prepare the scattering law in ENDF-6 format, there are a few hard coded options for the coherent elastic component, given by the \texttt{iel} flag in the input deck: \texttt{iel = 0} for no coherent elastic contribution, and options \texttt{iel = 1-6} for hard coded materials (Table \ref{tab:iel_leapr}). For the implementation of the mixed-elastic proposal, a new option (\texttt{iel = 100}, hereinafter referred to as MEF option) was added, which calculates and stores the parameters for the coherent and incoherent elastic components, as described in the proposed format. Since this change has not yet been widely adopted, an additional (\texttt{iel = 99}) option is provided to control the output of elastic data in the tsl-ENDF file using the current ENDF format (CEF), and hereinafter \texttt{iel = 99} option will be referred to as CEF option.

\begin{table}[h!]
\caption{Elastic options in LEAPR.}
\label{tab:iel_leapr}
\centering
\begin{tabular}{ c l }
\toprule
 \texttt{iel} & Description \\ 
 \midrule
0 & No coherent calculation.  \\
1 & Graphite (hard coded in LEAPR).  \\
2 & Beryllium (hard coded in LEAPR).  \\
3 & Beryllium oxide (hard coded in LEAPR).  \\
4 & Aluminum (hard coded in LEAPR).  \\
5 & Lead (hard coded in LEAPR).  \\
6 & Iron (hard coded in LEAPR).  \\
$\dots$ & Reserved.  \\
\color{red}\textbf{99} & Compute TSL in the current ENDF format (CEF).  \\
\color{red}\textbf{100} & Compute TSL in the proposed mixed elastic format (MEF).  \\
\bottomrule
\end{tabular}
\end{table}

\subsubsection{Mixed elastic format, MEF}

\begin{figure}[h!] 
  \centering
  \includegraphics[width=0.4\textwidth]{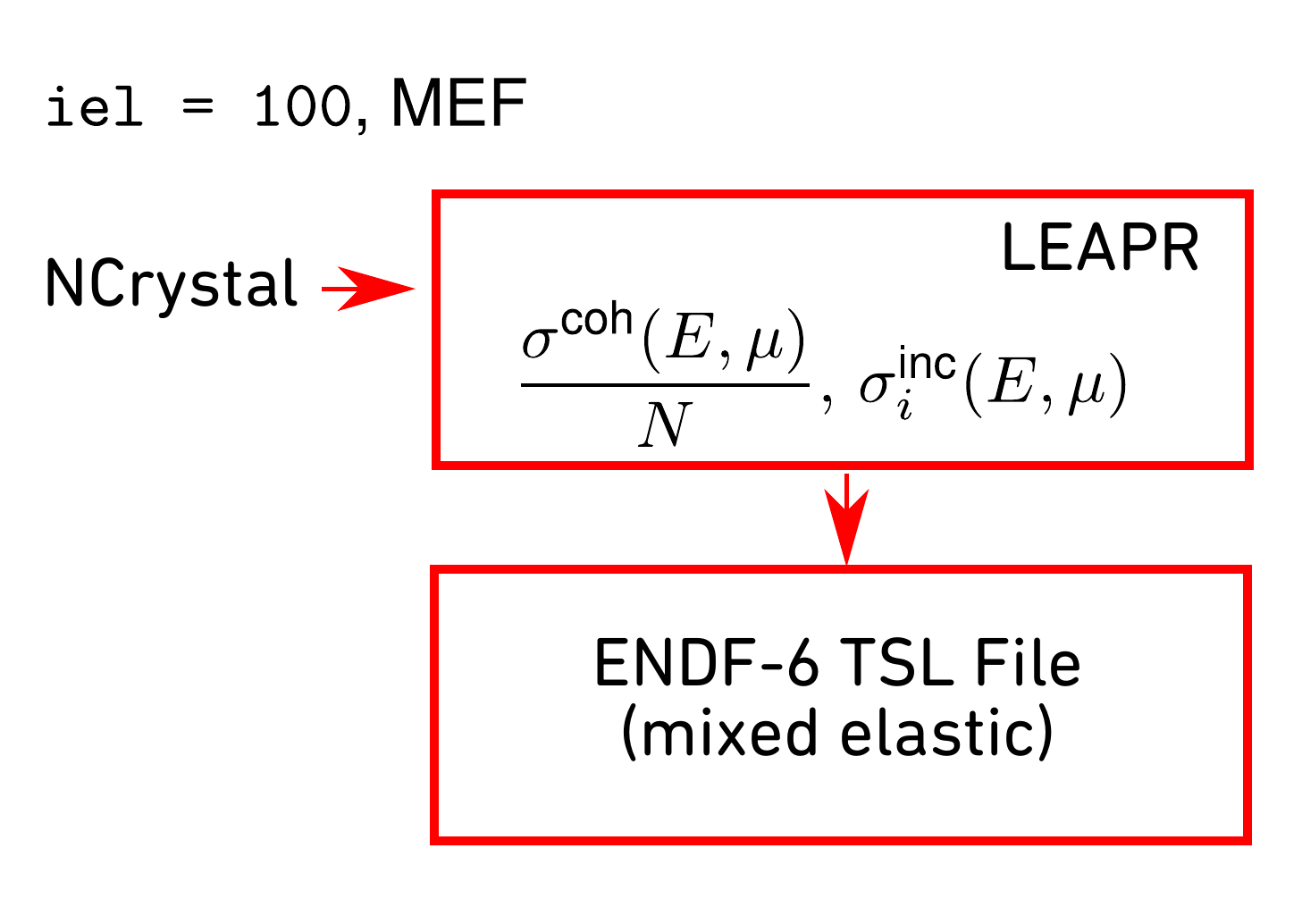}
  \caption{Simplified flow of NJOY+NCrystal when the MEF option is used.}
  \label{fig:elastic_options1}
\end{figure}

Coherent scattering is a property of the system (Equation \ref{eq:sqw_coh}), whereas incoherent scattering can be assigned individually to atoms (Equation \ref{eq:sqw_incoh}). In the ENDF-6 format, the scattering has to be decomposed into the contribution from the different elements. With this in mind, in the MEF option, the thermal scattering library for each element that composes a material contains two parts: coherent scattering divided by the total number of atoms in the system, and the incoherent scattering that corresponds to the atom (as can be seen in Figure \ref{fig:elastic_options1}):
\begin{equation}
    \sigma^\text{el}_i(E, \mu) = \sigma^\text{coh}(E, \mu)/N +  \sigma^\text{inc}_i(E, \mu).
\end{equation}

\subsubsection{Current ENDF format, CEF}

\begin{figure}[h!] 
  \centering
  \includegraphics[width=0.6\textwidth]{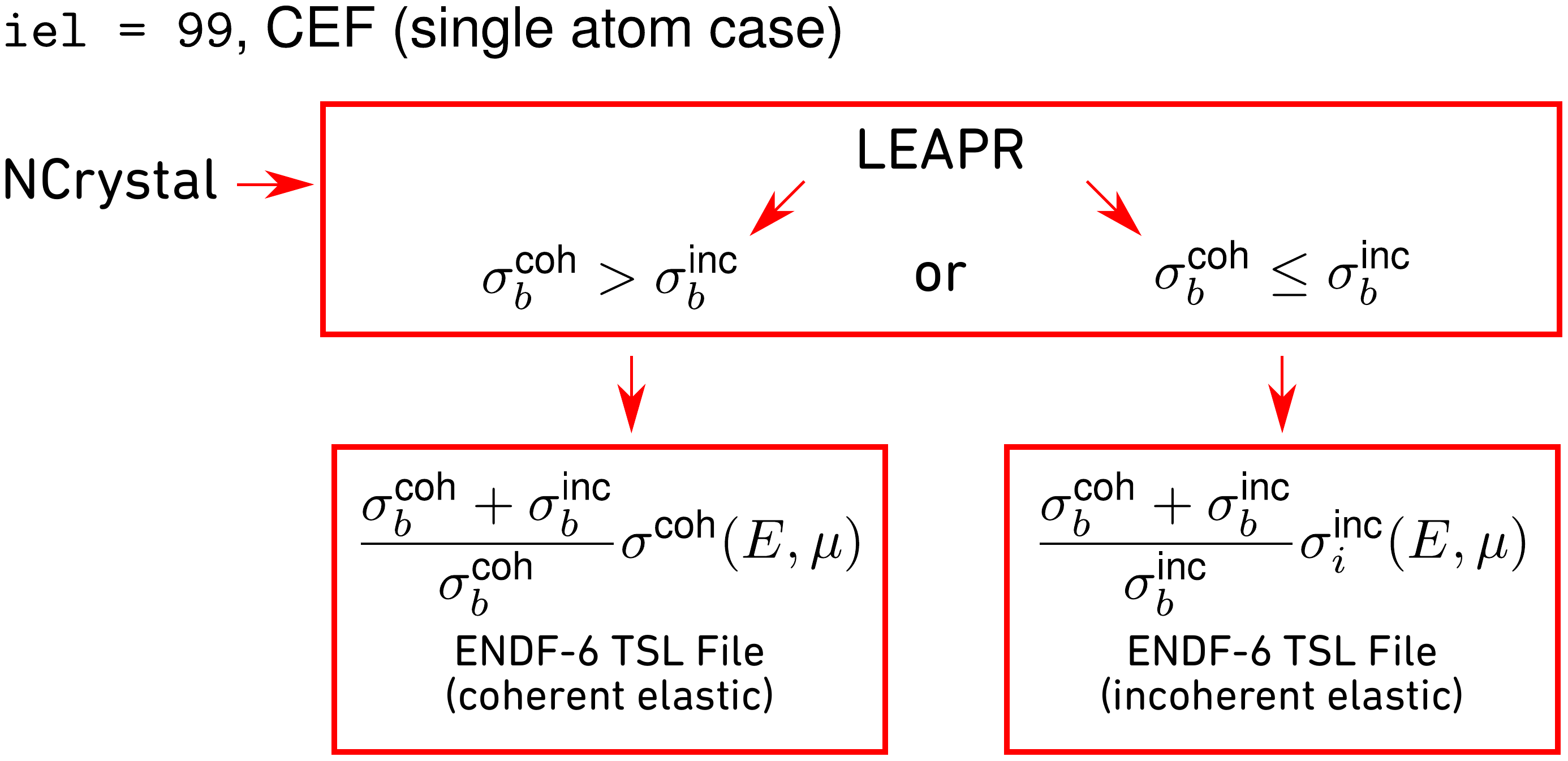}
  \caption{Simplified flow of NJOY+NCrystal when the CEF option is used for a monatomic scatterer.}
  \label{fig:elastic_options2}
\end{figure}

In the CEF option, only the coherent or incoherent scattering model can be used, but not both. For systems with only one type of atom (e.g. Al, Fe, Si), the major component is stored but it is scaled to the total bound scattering cross section (as can be seen in Figure \ref{fig:elastic_options2}) to ensure the proper asymptotic behavior in the epithermal range:

\begin{equation}
    \sigma^\text{el}_i(E, \mu) = \dfrac{\sigma_b^\text{coh}+\sigma_b^\text{inc}}{\sigma_b^\text{coh}}\sigma^\text{coh}(E, \mu),\;\; \text{if}\; \sigma_b^\text{coh} > \sigma_b^\text{inc},\; \text{(coherent approximation)} \label{eq:cef1}
\end{equation}

\begin{equation}
    \sigma^\text{el}_i(E, \mu) = \dfrac{\sigma_b^\text{coh}+\sigma_b^\text{inc}}{\sigma_b^\text{inc}} \sigma^\text{inc}_i(E, \mu),\;\; \text{if}\; \sigma_b^\text{coh} \leq \sigma_b^\text{inc},\; \text{(incoherent approximation)}\label{eq:cef2}
\end{equation}

\begin{figure}[h!] 
  \centering
  \includegraphics[width=0.6\textwidth]{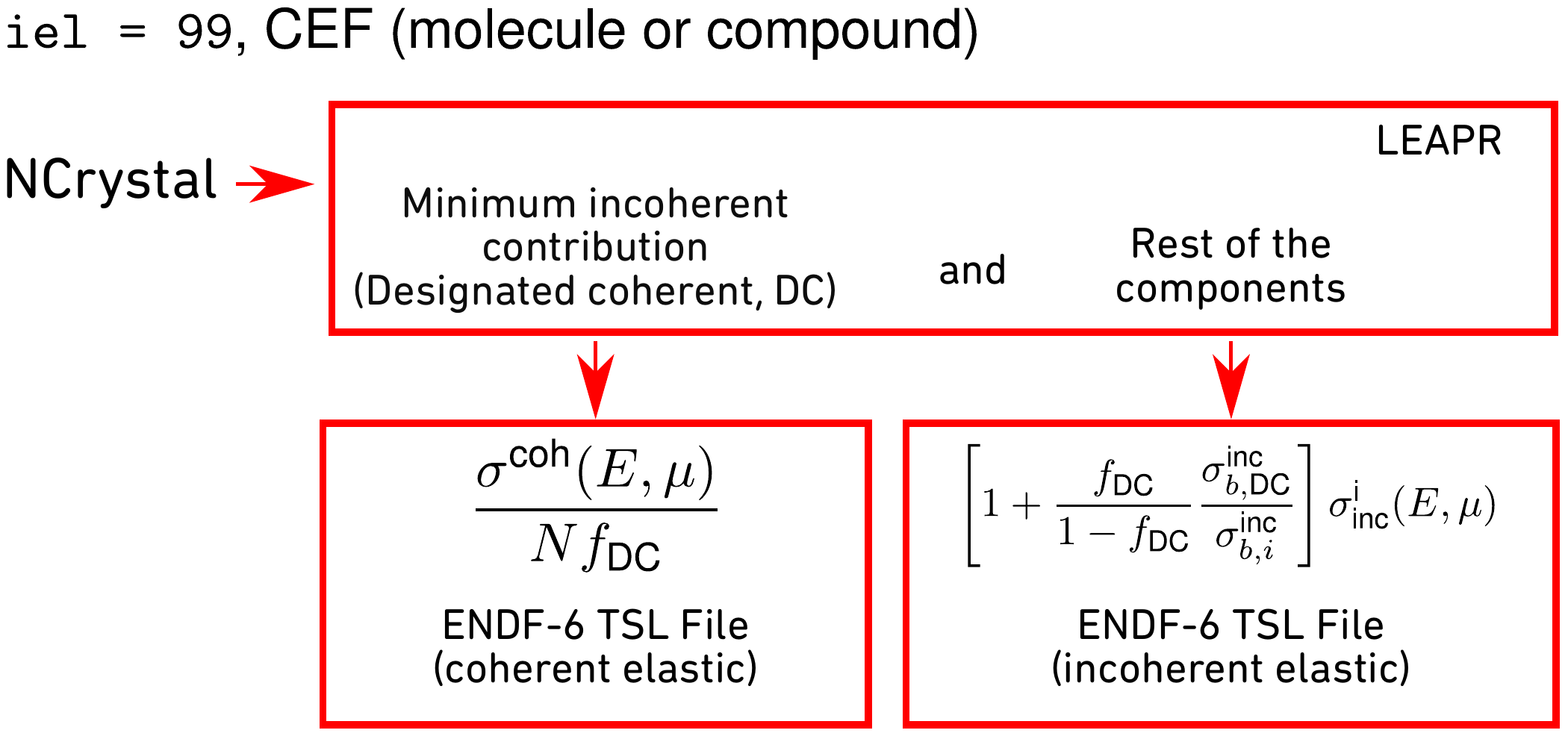}
   \caption{Simplified flow of NJOY+NCrystal when the CEF option is used for a polyatomic or molecular scatterer.}
   \label{fig:elastic_options3}
\end{figure}

For systems with more than one atom, the contributions are sorted and the atom with the minimum incoherent contribution (called \emph{designated coherent}, DC) is stored in the coherent approximation. Its incoherent contribution, if it exists, is distributed among the rest of the elements which are stored in the incoherent approximation (as can be seen in Figure \ref{fig:elastic_options3}):

\begin{equation*}
    \sigma^\text{el}_\text{mol}(E, \mu) = N \left[\sigma^\text{coh}(E, \mu)/N \right] + N_{DC} \sigma^\text{DC}_\text{inc}(E,\mu) + \sum_{i\neq DC} N_i \sigma^i_\text{inc}(E,\mu)
\end{equation*}

\begin{equation*}
     = N_\text{DC} \left[\sigma^\text{coh}(E, \mu)/(N f_\text{DC}) \right] + \sum_{i\neq DC} N_i \left[\sigma^i_\text{inc}(E,\mu) + \dfrac{f_\text{DC}}{1-f_\text{DC}}\sigma^\text{DC}_\text{inc}(E,\mu)\right] 
\end{equation*}

\begin{equation}
     \approx N_\text{DC} \left[\sigma^\text{coh}(E, \mu)/(N f_\text{DC}) \right] + \sum_{i\neq DC} N_i \left[1 + \dfrac{f_\text{DC}}{1-f_\text{DC}} \dfrac{\sigma^\text{inc}_{b,\text{DC}}}{\sigma^\text{inc}_{b,i}}\right] \sigma^\text{i}_\text{inc}(E,\mu)\label{eq:cef3}
\end{equation}

The first term in this equation is stored in the elastic component of the DC atom in the coherent approximation, and each term of the sum in the second term is stored in the rest of the atoms in the incoherent approximation. Since incoherent scattering of the designated coherent element is being redistributed between the other atoms in the molecule/compound, the approximation made in the last step impĺies using the Debye-Waller factor of the rest of the components to account for the missing incoherent scattering of the designated coherent element.

If the mixed-elastic proposal cannot be used, the calculation of the CEF option using NJOY+NCrystal provides a reasonable approximation within the limitations of the format. It is important to note that the CEF option is currently supported in standard versions of MCNP and PHITS, whereas the MEF option requires modifications. Examples of this are provided in Section \ref{sec:validation_and_examples}. It is also important to note that both the CEF and the MEF options will only provide the right answer if they are used in the correct stoichiometry.

\subsubsection{Wrapper function}

For the CEF and MEF options, LEAPR calls a wrapper function (since NJOY2016 is written in Fortran while NCrystal is in C++ programing language) that for the coherent elastic component obtains a list of Bragg edges, given as pairs of energy and cross section data for each Bragg edge from NCrystal. The wrapper function also passes to LEAPR the values for the bound coherent and incoherent elastic cross sections, as well as the free atom cross section for the nuclide that is being processed in LEAPR, which is used for consistency between the inelastic and elastic components. These cross sections are obtained from a builtin library inside NCrystal, which includes data for all the major elements and isotopes. 
Since the elastic component is handled by NCrystal and the inelastic by LEAPR module of NJOY+NCrystal, an additional consistency check is performed by comparing the mean squared displacement (MSD) values from the two codes for the nuclide that is being processed in LEAPR. The wrapper function passes to LEAPR the NCrystal MSD value and both NCystal and NJOY+NCrystal values for MSD are stored in the output file of the LEAPR module. Since both the inelastic and elastic component exhibit dependence on the phonon spectrum (for the elastic component the dependence is through Debye-Waller integral as seen in Equations \ref{eq:isotropic_debye_waller_factor},\ref{eq:thermr_ncrystal_dw},\ref{eq:thermr_dw_phonon_dos}, and for the inelastic dependence can be seen in Equations \ref{eq:SAlphaAndBeta},\ref{eq:gammat},\ref{eq:PBeta}), the check of the MSD values is also a check on whether the same phonon spectrum is used in both the NCrystal and NJOY+NCrystal input files. The LEAPR value of MSD can be calculated from $W_\text{ENDF}$, as defined in Equation \ref{eq:thermr_dw_phonon_dos}, as follows:
\begin{equation}
    <u^2>=\frac{\hbar^2W_\text{ENDF}}{2m_n},
\end{equation}
where $m_n$ is the mass of neutron in units of $\unit{eV}\cdot \unit{ps}^2\cdot\SI{}{\angstrom}^{-2}$, and $\hbar^2$ is Planck's constant in units of $\unit{eV}\cdot \unit{ps}$.

\subsection{THERMR/ACER modifications}\label{therm_acer_modifications}
The modules THERMR and ACER of NJOY were also modified to implement processing of the tsl-ENDF proposed mixed elastic format, as well as to produce .ACE files in a format that is being proposed here, since at the time of the development of the code no official format for mixed elastic implementation in .ACE files has been developed or announced. THERMR reconstructs the double differential cross section and produces pointwise ENDF files (PENDF tapes) from tsl-ENDF files, and ACER processes PENDF files into .ACE files.

THERMR reads the tsl-ENDF file and handles the elastic section according to the \texttt{lthr} flag: \texttt{lthr=1} for coherent elastic, \texttt{lthr=2} for incoherent elastic or here proposed \texttt{lthr=3} for mixed elastic. There are no changes to the THERMR input deck and the \texttt{lthr} flag is set during the writing process of the tsl-ENDF file in LEAPR. 

\begin{figure}[t!] 
  \centering
  \includegraphics[width=1.00\textwidth]{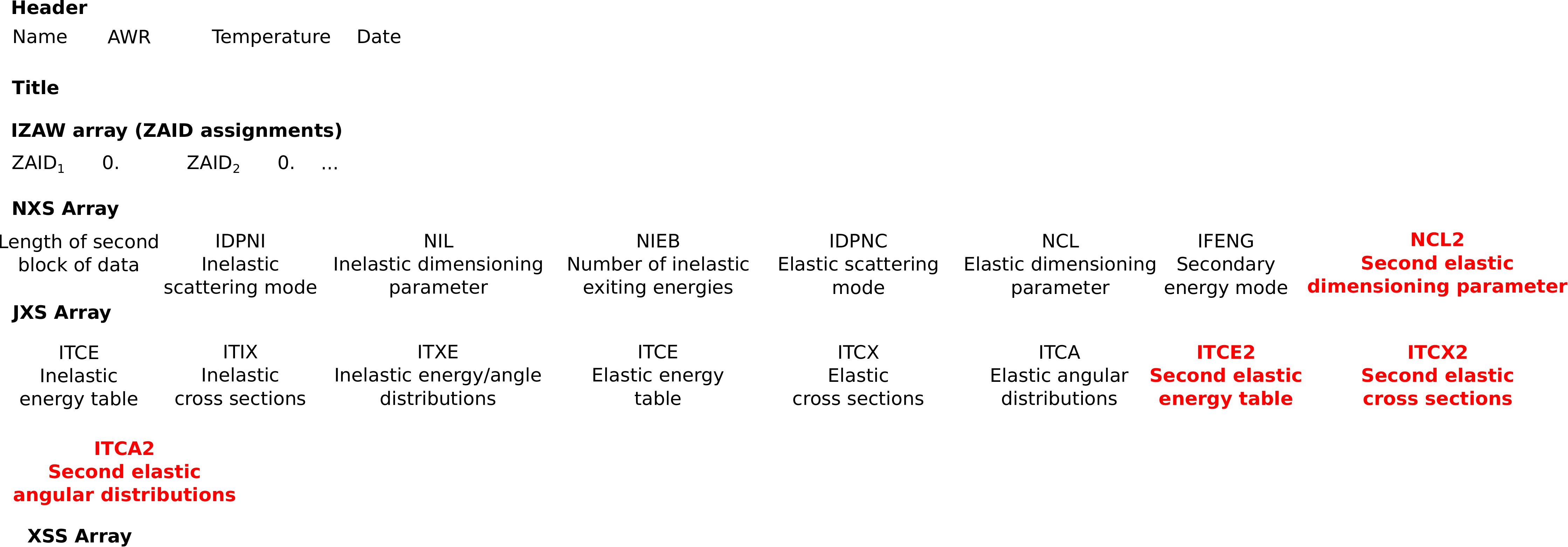}
  \caption{Proposed format for the thermal .ACE files with mixed elastic. Changes are highlighted in red.}
  \label{fig:ace_format_mixed_elastic}
\end{figure}

In the ACER input deck, an additional option for the \texttt{ielas} flag was added. Option \texttt{ielas=0} processes the PENDF tape as coherent elastic, \texttt{ielas=1} processes the file as incoherent elastic, and the here proposed \texttt{ielas=2} as mixed elastic. The .ACE format \cite{ref_ace} requires changes to store the additional data for the mixed elastic contributions, and the changes we propose in this work can be seen in Figure \ref{fig:ace_format_mixed_elastic}. In the proposed format, the flag \texttt{IDPNC=5} is reserved for mixed elastic, and four additional integers are stored in the NXS and JXS control arrays. \texttt{NCL2} determines the dimension of the second elastic block, \texttt{ITCE2} gives the position of the energy table for the second elastic block, \texttt{ITCX2} gives the position of the cross section table for the second elastic block, and \texttt{ITCA2} gives the position of the angular distribution. In this implementation of the mixed elastic format of .ACE files, the first elastic block is coherent and the second elastic block is an incoherent elastic component.

\subsection{OpenMC modifications}\label{openmc_modifications}

As a proof of concept implementation we modified OpenMC to read and use .ACE files in the proposed mixed elastic format. The Python API was modified to read mixed elastic .ACE files and as well to read and write HDF5 files in mixed elastic format. The C++ calculation engine was modified to add the second elastic component to the total cross section if it is present and to sample both elastic reactions. The modifications to OpenMC can be found at \cite{wwwhighnessopenmc}.

\section{Example evaluations}\label{sec:validation_and_examples}


Using NJOY+NCrystal, 213 tsl-ENDF evaluations were created for 112 new or updated materials. A summary of the NJOY+NCrystal library is given in Tables \ref{table:endf_libraries}-\ref{table:endf_libraries5}.  In this section, a few materials will be highlighted, which include tin as an example of a coherent scatterer, vanadium as an incoherent scatterer, nickel as mixed elastic scatterer, and MgH$_2$ and MgD$_2$ are shown as relevant materials to the HighNESS project.

For the materials in Tables \ref{table:endf_libraries}-\ref{table:endf_libraries5}, different methodologies were employed to create the final tsl-ENDF files. Some of the materials were created from existing .NCMAT files in current and previous versions of NCrystal. For some of the materials, mostly monoatomic metals, VDOS curves in existing .NCMAT files were updated, mostly using different sources from the literature. For MgH$_2$ and MgD$_2$, detailed DFT calculations were performed and the details can be found in Section \ref{subsection:mgh2_and_mgd2}. For the rest of the materials, phonon curves were obtained by utilizing ab-initio calculations from a phonon database at Kyoto University \cite{wwwphonodb,doi:10.1063/1.4812323,ONG2013314,ONG2015209}. In summary, the files from the database were used with Phonopy \cite{ref_phonopy} to calculate phonon eigenvalues and eigenfrequencies, which were then utilized with oClimax \cite{ref_oclimax} to extract partial phonon spectra. A detailed explanation of how the phonon spectrum and crystal structure was obtained for each material is provided in the comments section of the tsl-ENDF files. 

The experimental data for the validation of the tsl-ENDF files is scarce. Wherever the experimental data were available, tsl-ENDF files were validated against total cross section measurements and diffraction data, while all materials were validated against specific heat capacity curves as a minimum standard for acceptance. The last column of Tables \ref{table:endf_libraries}-\ref{table:endf_libraries5} contains information on the valid temperature range for each material, references for VDOS curves, and a list of validations applied. For the VDOS curves a reference is provided, where for the ones calculated with Phonopy and oClimax the reference to Kyoto University phonon database is given. The link to the exact files used for each material at Kyoto University database can be found inside the tsl-ENDF files in the Highness Github repository. For validation, the correct crystal structure was a starting point (a reference for each crystal structure can be found inside tsl-ENDF files as well), followed by comparison with the experimental specific heat capacity and if available, total cross section measurements. For $\alpha$ and $\beta$ SiC, diffraction data was used as well as a means of validation of the libraries.


\subsection{Tin}
\begin{figure}[h!] 
  \centering
  \includegraphics[scale=0.5]{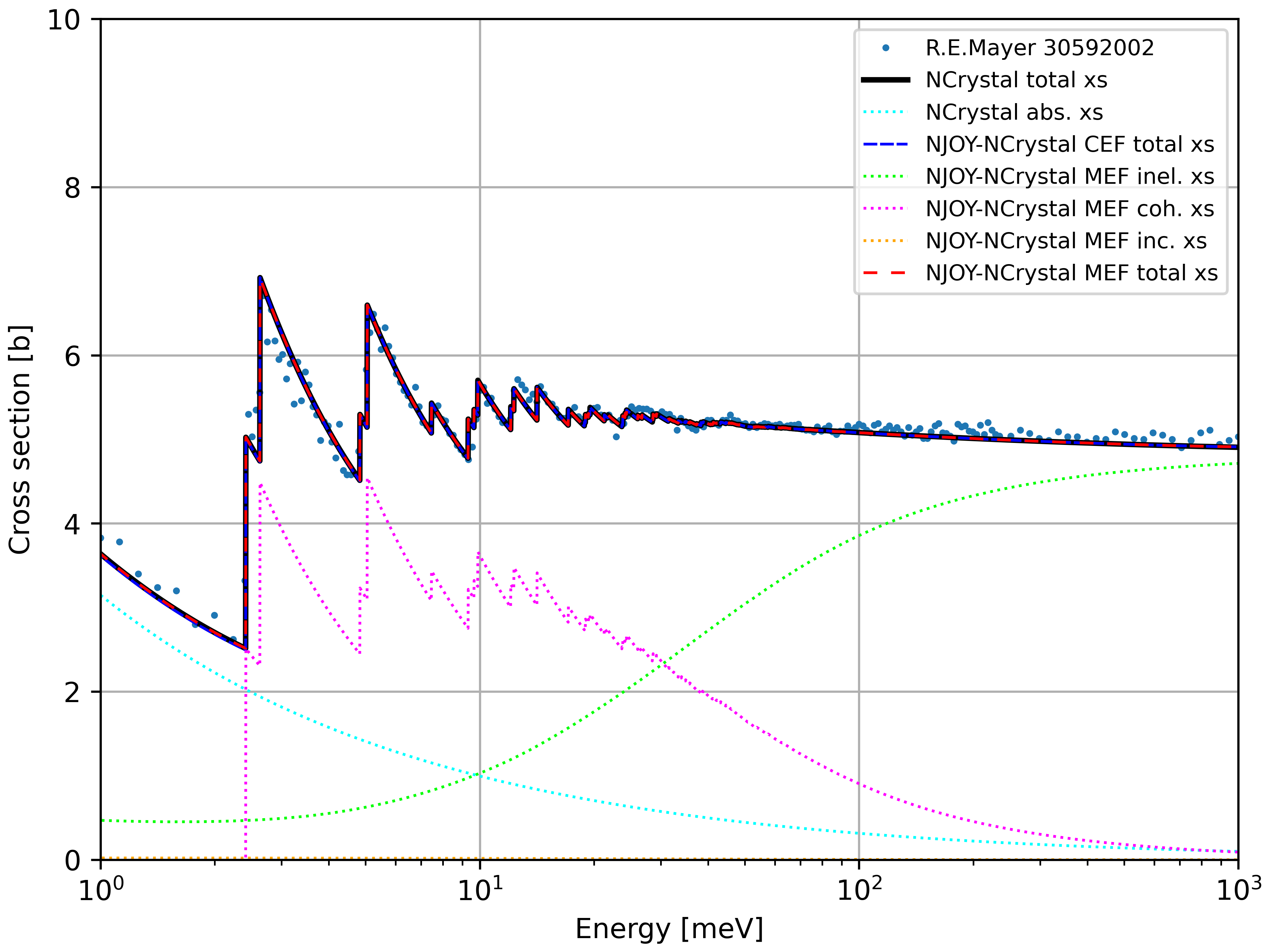}
  \caption{Total scattering cross section comparison for tin.}
  \label{fig:tin_total_xs}
\end{figure}

With $\sigma_\text{inc} = 0.022$ b and $\sigma_\text{coh}=4.870$ b, tin is a mainly coherent scatterer. Using NJOY+ NCrystal we have created tsl-ENDF and .ACE files for tin, and compared the calculated total cross section with measurements by Mayer\cite{mayer1981total,mayer1981total_tin-exfor}, retrieved from EXFOR\cite{otuka2014towards} (Figure \ref{fig:tin_total_xs}). In this comparison, both scattering and absorption are included, using $\sigma_{2200} = 0.626$ b. For completeness, the total cross section from the .ACE files, prepared by NJOY+ NCrystal, was plotted using OpenMC and the results for CEF and MEF options are shown, as well as each scattering component from the MEF option. 

In Figure \ref{fig:tin_total_xs} it can be observed that the agreement between the experimental total cross sections and the calculated cross sections for NCrystal, and using both CEF and MEF options in NJOY+NCrystal is very good (all three lines are overlaid on top of each other). Therefore, for predominantly coherent materials like tin, the CEF option performs nearly as well as the MEF option, and it can be used even without the mixed-elastic format.

\subsection{Vanadium}
Vanadium is a common example of a mostly incoherent material, with $\sigma_\text{inc} = 5.08$ b and $\sigma_\text{coh} = 0.018$ b. The calculated cross sections were obtained in the same manner as for tin.

\begin{figure}[h!] 
  \centering
  \includegraphics[scale=0.5]{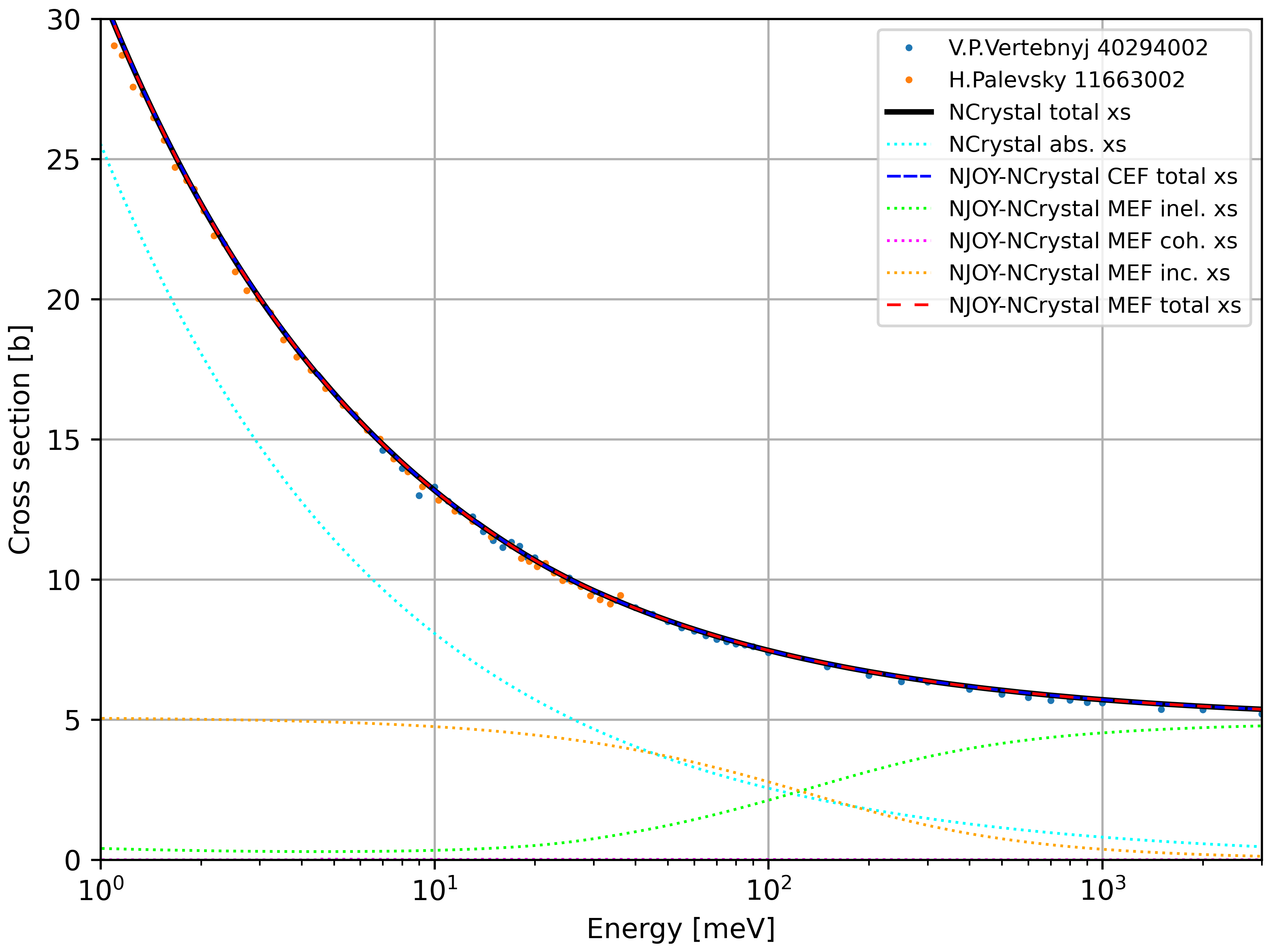}
  \caption{Total scattering cross section comparison for vanadium.}
  \label{fig:vanadium_total_xs}
\end{figure}

As can be seen from Figure \ref{fig:vanadium_total_xs}, the calculated total cross sections, including $\sigma_{2200} = 5.08$ b, are in excellent agreement with the experimental total cross section measurements from Vertebnyj \cite{vertebnyi_vanadium-exfor} and Palevsky \cite{palevsky_vanadium-exfor}. The coherent elastic component is negligible, while the incoherent elastic component is dominant. Again, the CEF option provides an adequate representation of the total cross section for Vanadium.

\subsection{Nickel}
Nickel, with $\sigma_\text{inc} = 5.2$ b and $\sigma_\text{coh} = 13.3$ b, has a significant contribution from both coherent and incoherent scattering. This, in the current ENDF-6 format, can only be approximated correctly to a certain degree.
\begin{figure}[h!] 
  \centering
  \includegraphics[scale=0.5]{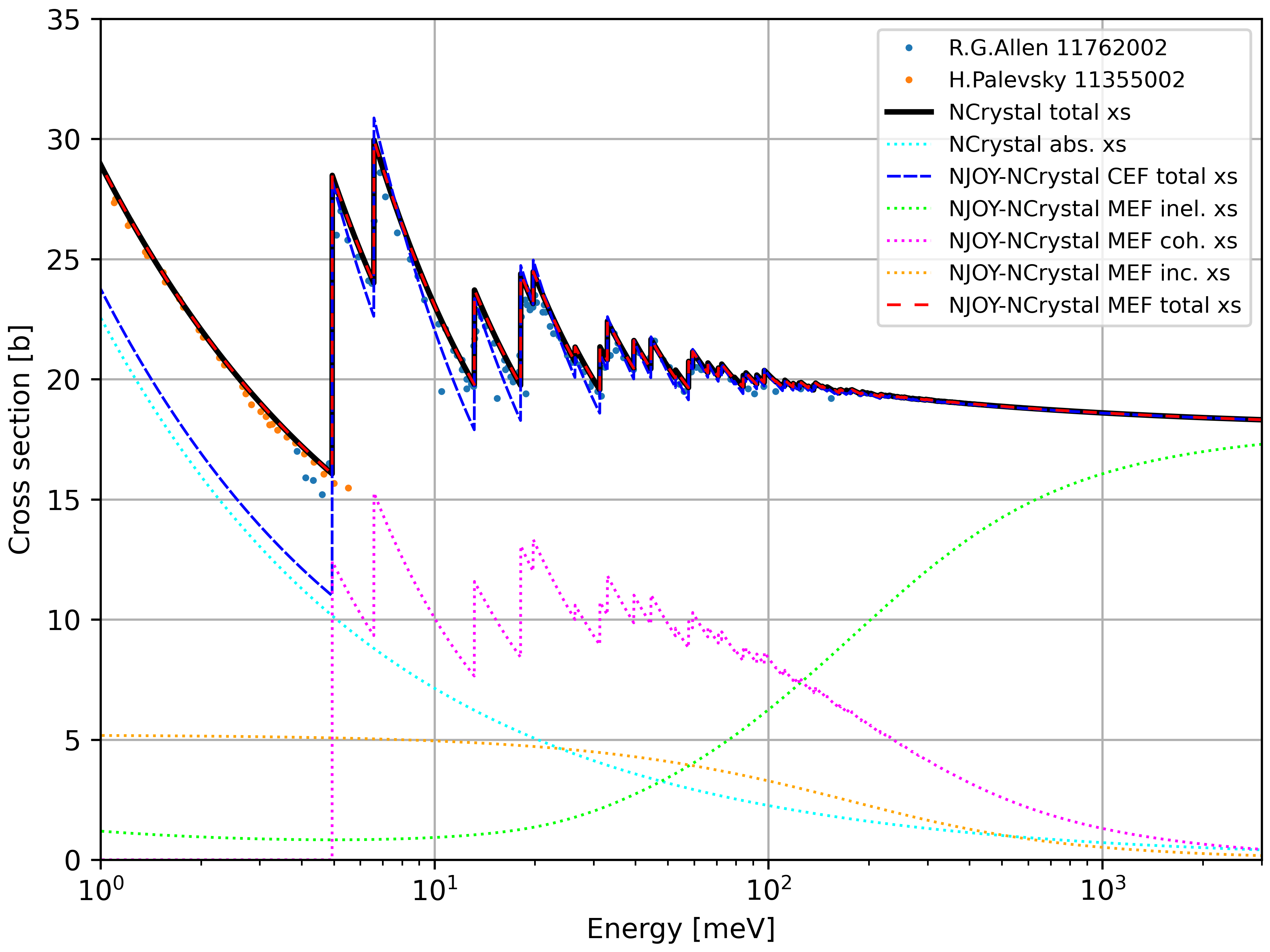}
  \caption{Total scattering cross section comparison for nickel.}
  \label{fig:nickel_total_xs}
\end{figure}

In Figure \ref{fig:nickel_total_xs} we compare the calculated total cross sections, including $\sigma_\text{2200} = 4.49$ b, with experimental data from Allen \cite{allen1954slow,allen1954slow-exfor} and Palevsky \cite{hughes1958neutron, hughes1958neutron-exfor}. In this plot it can be seen that the agreement of the NCrystal and the total cross section curve calculated with the MEF option with the experimental data is excellent, while the agreement for the CEF option curve is good in the region above the first Bragg edge. Since the CEF option scales the major elastic component, in this case the coherent elastic part (Equation \ref{eq:cef1}), by the total bound incoherent plus coherent scattering cross section, the total cross section below the first Bragg edge only contains the contribution from the inelastic component and hence differs significantly. There is also some discrepancy above the first Bragg edge as well for CEF option, which is due to the fact that coherent elastic scaling provides the correct epithermal asymptotic behavior but only approximates the low energy region. Thus this demonstrates the importance for the ENDF-6 format to support the capability to store both the coherent and incoherent elastic scattering components.

\begin{figure}[h!] 
  \centering
  \includegraphics[scale=0.5]{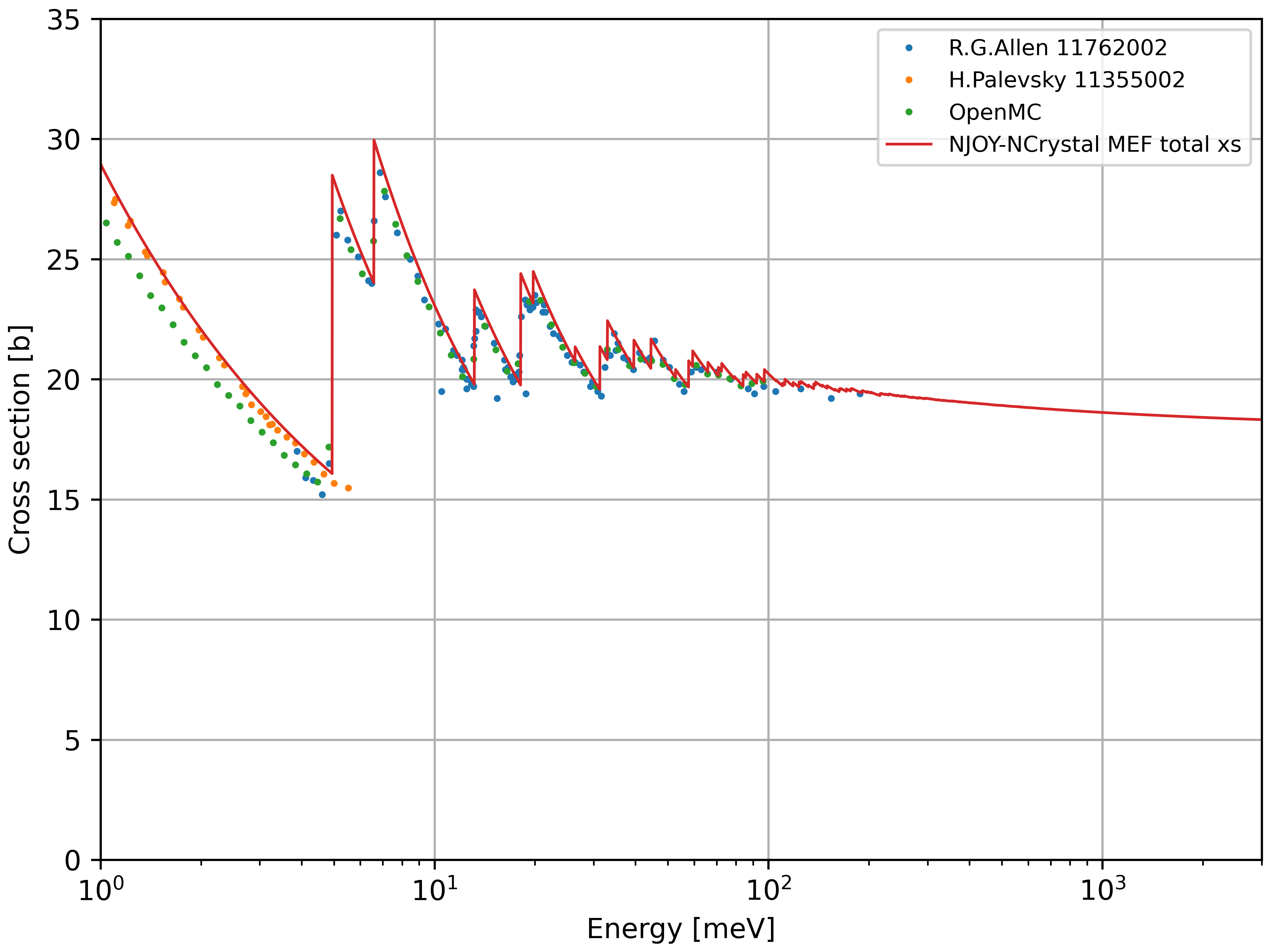}
  \caption{OpenMC calculation of the simplified transmission experiment on nickel. The difference in the low energy region is caused by differences in the absorption cross section. NCrystal uses $\sigma_{2200}$ = 4.49 barns whereas in ENDF/B-VIII.0 $\sigma_{2200}$ = 4.09 barns.}
  \label{fig:nickel_openmc_total_xs}
\end{figure}

As mentioned in Section \ref{openmc_modifications}, changes to OpenMC were implemented so that the new format can be used to sample thermal scattering events. Figure \ref{fig:nickel_openmc_total_xs} shows the results of a simple transmission Monte Carlo calculation, which includes a mono-energetic neutron beam on a 5 mm thick slab of nickel. The total cross section curve was obtained from the ratio of the incident and the transmitted neutron spectrum and we can see that the agreement between the calculation and the experimental data,  as well as the total cross section calculated in the MEF option, is excellent. The discrepancy at the lower energies is caused by differences in the absorption cross section in ENDF/B-VIII.0 library ($\sigma_{2200}= 4.09$ b) and NCrystal ($\sigma_{2200}= 4.49$ b). 

\subsection{MgD$_{2}$/MgH$_{2}$}\label{subsection:mgh2_and_mgd2}

As an additional example of the capabilities of NJOY+ NCrystal, related to the HighNESS project, we computed the total scattering neutron cross sections of polycrystalline MgH$_2$ and its deuterated variant MgD$_2$. MgH$_2$ recently emerged as a good candidate material for cold neutron reflectors \cite{granada2021development},  while MgD$_2$, because of an even lower absorption cross section, is also investigated here. 

Phonon spectrum calculations were performed using Density Functional Perturbation Theory (DFPT) \cite{Baroni} with the Perdew-Burke-Ernzerhof (PBE) \cite{pbe} approximation to the exchange and correlation energy functional. Ultrasoft pseudopotentials \cite{uspseudo,Dalcorso2014} and a plane wave expansion of Kohn-Sham orbitals up to a kinetic cutoff of 60 Ry was employed, as implemented in the Quantum-ESPRESSO package \cite{QMEspresso}. In the electronic structure calculations, the Brillouin zone (BZ) integration was performed over a uniform $\Gamma$-centered 10$\times$10$\times$14 \textbf{k}-point mesh \cite{mp}.

MgH$_2$ crystallizes in the rutile structure (P4$_2$/mmm space group) with six atoms per unit cell \cite{Ellinger1955}. Geometry optimization yields the theoretical equilibrium lattice parameters a=4.512 \AA\ and c=3.010 \AA\ which are very close to the experimental values of a=4.5025 \AA\ and c=3.0123 \AA\ as measured from X-ray scattering \cite{Ellinger1955}. The dynamical matrix is computed within DFPT on a 4$\times$4$\times$6 \textbf{q}-mesh. Short range force constants were then obtained by Fourier transforming the dynamical matrices over the \textbf{q}-mesh after subtracting the contribution from point charges as described in \cite{Baroni}. The force constants then allow computing the phonon spectrum on a fine \textbf{q}-mesh \cite{Baroni}. 
 
\begin{figure}[h!]
    \includegraphics[width=0.49\columnwidth]{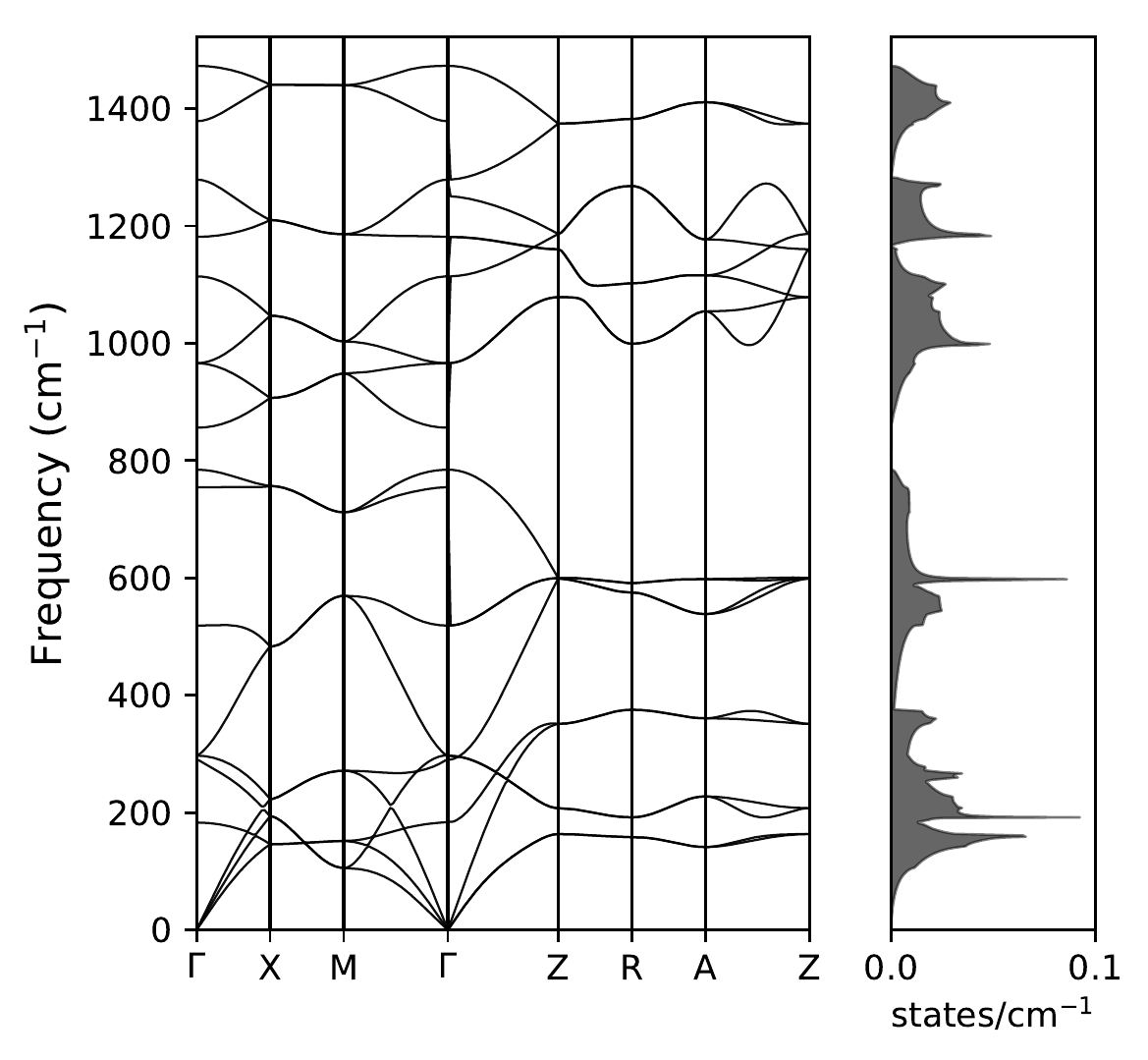}
    \includegraphics[width=0.49\columnwidth]{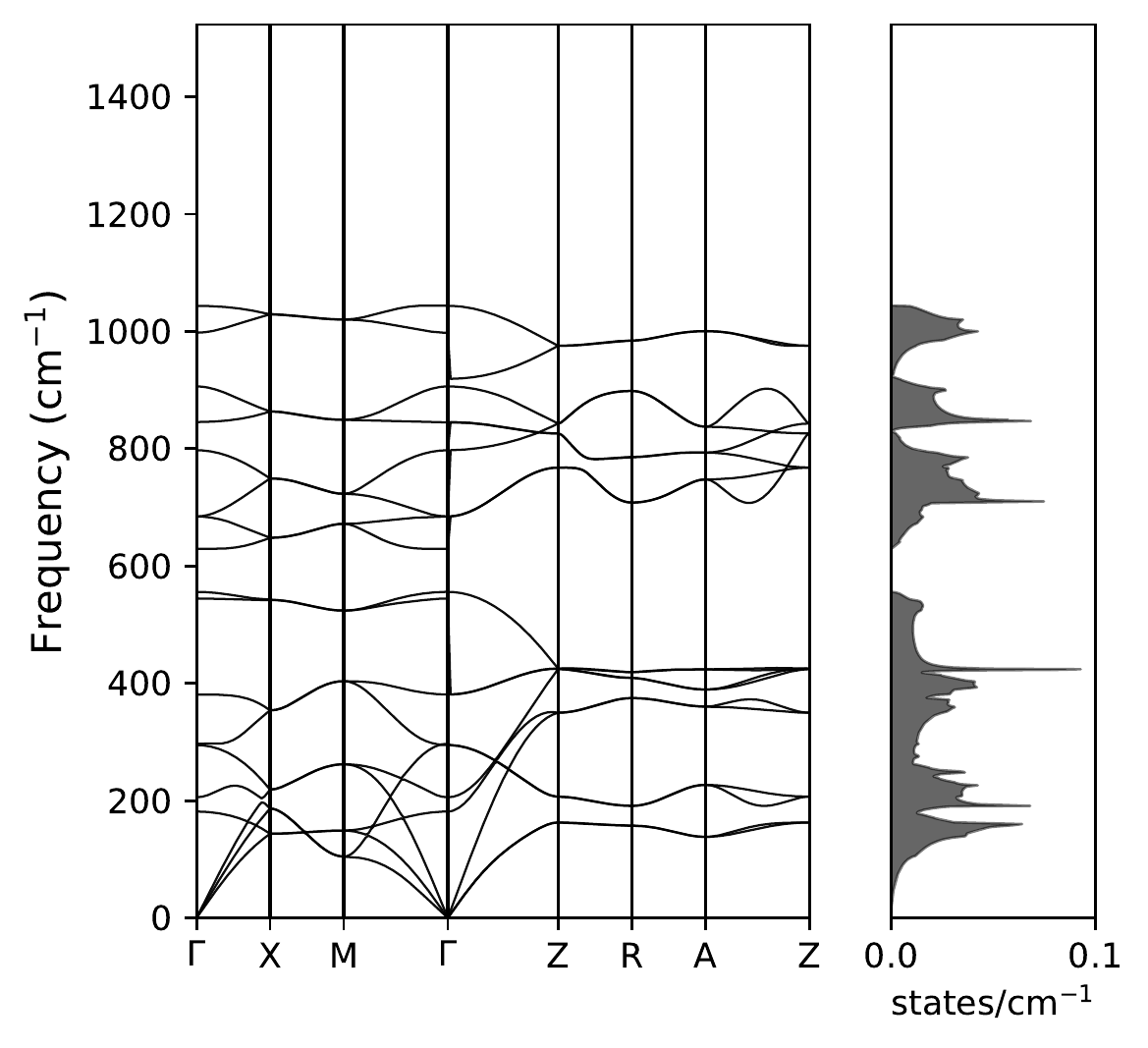}
\caption{Theoretical phonon dispersion relations along high-symmetry directions in the Brillouin zone and total phonon density of states for (left) MgH$_2$ and (right) MgD$_2$.}
\label{fig:H2D2comp}
\end{figure}

\begin{figure}
	\centering
	\includegraphics[scale=0.5]{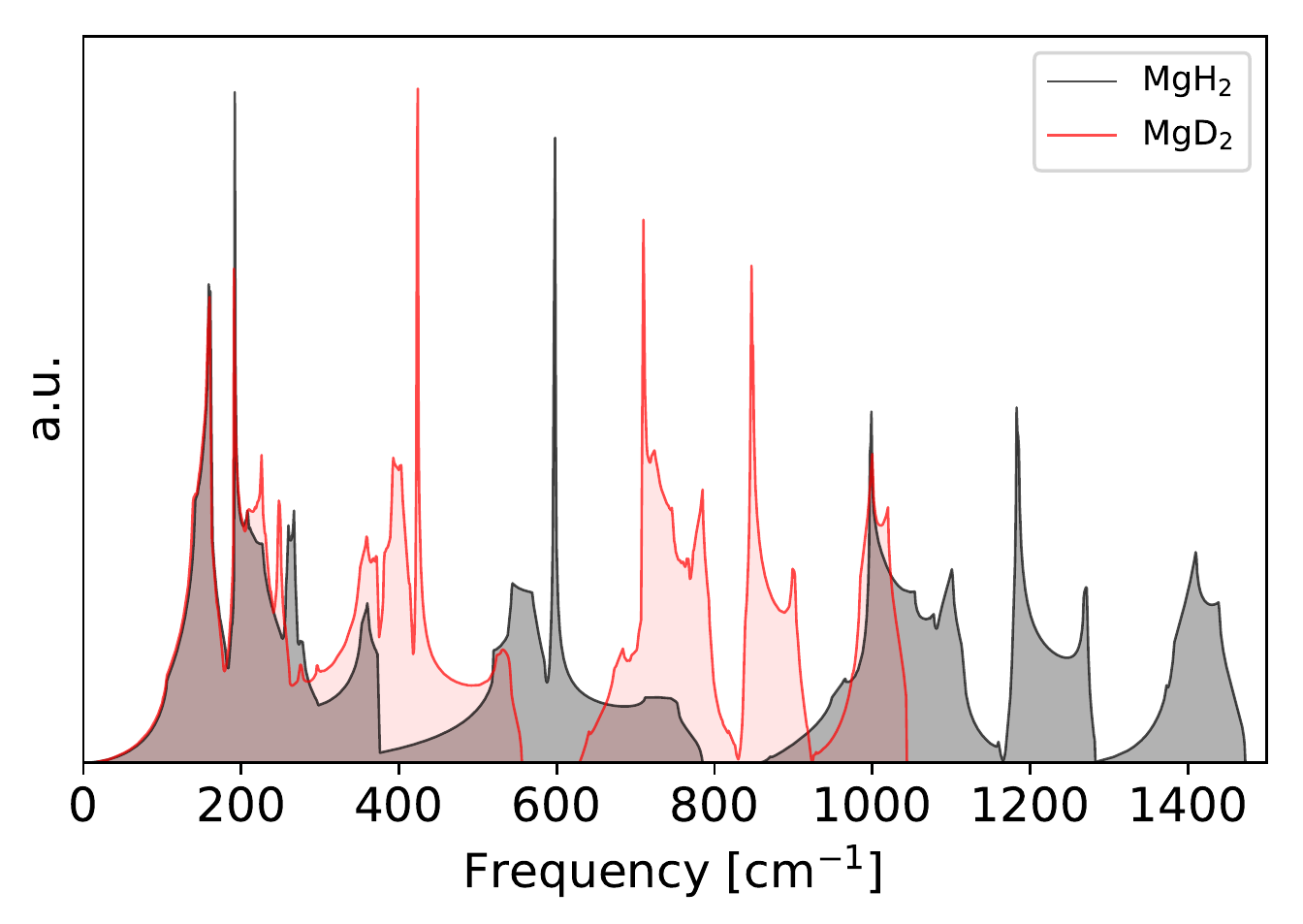}
	\caption{Comparison of the theoretical phonon DOS for MgH$_2$ and MgD$_2$. The density of states have been computed using the tetrahedron method using a 20$\times$20$\times$30 \textbf{q}-grid for interpolation.}
	\label{fig:phdos}
\end{figure}

The phonon dispersion relations and the DOS of MgH$_2$ and MgD$_2$ are compared in Figures \ref{fig:H2D2comp}-\ref{fig:phdos}. The results for MgH$_2$ are very similar to previous ab-initio works \cite{Ohba,Schimmel}. In particular, in \cite{Ohba} the overall validity of the isotropic approximation of the mean square displacement is assessed at different temperatures. The high-frequency phonons, due to the motion of hydrogen, are almost exactly scaled by a factor $\sqrt{2}$ by isotopic substitution while the acoustic part of the spectrum is essentially unchanged. 
  
\begin{figure}
	\centering
	\includegraphics[scale=0.5]{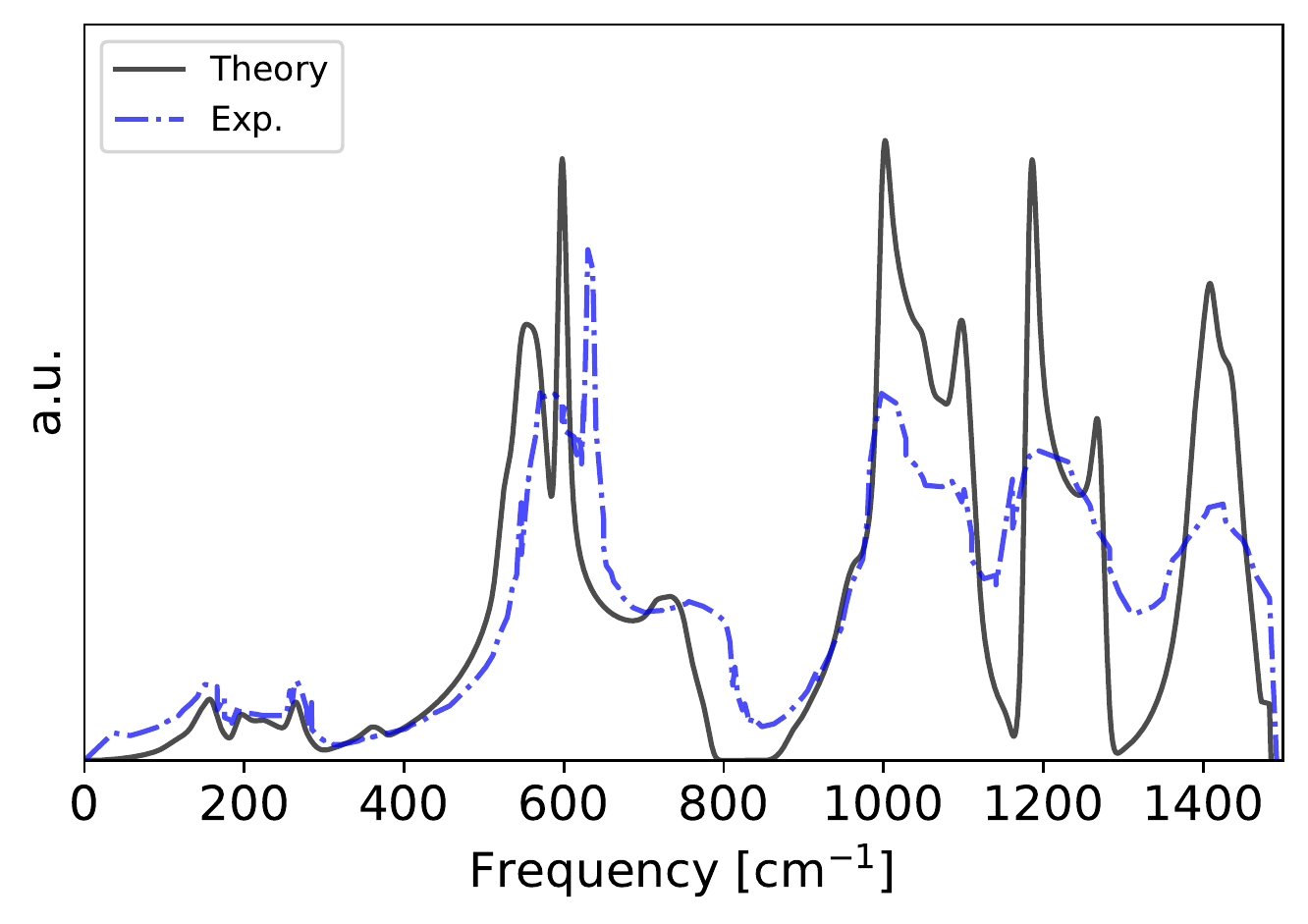}
	\caption{Neutron-weighted theoretical phonon DOS compared with an experimental spectrum from inelastic neutron scattering \cite{Kolesnikov2011}. The theoretical density of states has been computed using the tetrahedron method using a 20$\times$20$\times$30 \textbf{q}-grid for interpolation and then convoluted with a Gaussian function with standard deviation of 5 cm$^{-1}$ to better match the experimental resolution.}
	\label{fig:nxs_phdos}
\end{figure}

The neutron-weighted phonon density of states, computed by multiplying the projected DOS with the corresponding total neutron scattering cross section\cite{neutronxs}, is compared in Figure \ref{fig:nxs_phdos} with the experimental spectrum from inelastic neutron scattering data \cite{Kolesnikov2011}. The agreement between theory and experiment is overall good, although a sizable error is present in the position of the peak at about 600 cm$^{-1}$. A good agreement between theory and experiments is also found for the frequency of the Raman active modes ($\Gamma$ point) as shown in Table \ref{tab:raman}.

\newpage
\begin{table}[h]
\caption{Theoretical (this work) and experimental \protect\cite{Cai2018} frequency (cm$^{-1}$) of Raman active modes of MgH$_2$.}
\label{tab:raman}
\begin{tabular}{ c c c }
\toprule
 & Exp. & Theory \\ 
 \midrule
B$_{\rm 1g}$ & 302.65 & 290  \\
\hline
E$_{\rm g}$ & 967.36 & 966 \\
\hline
A$_{\rm 1g}$ & 1295.79 & 1279 \\
\bottomrule
\end{tabular}
\end{table}

\begin{figure}[h!] 
  \centering
  \includegraphics[scale=0.5]{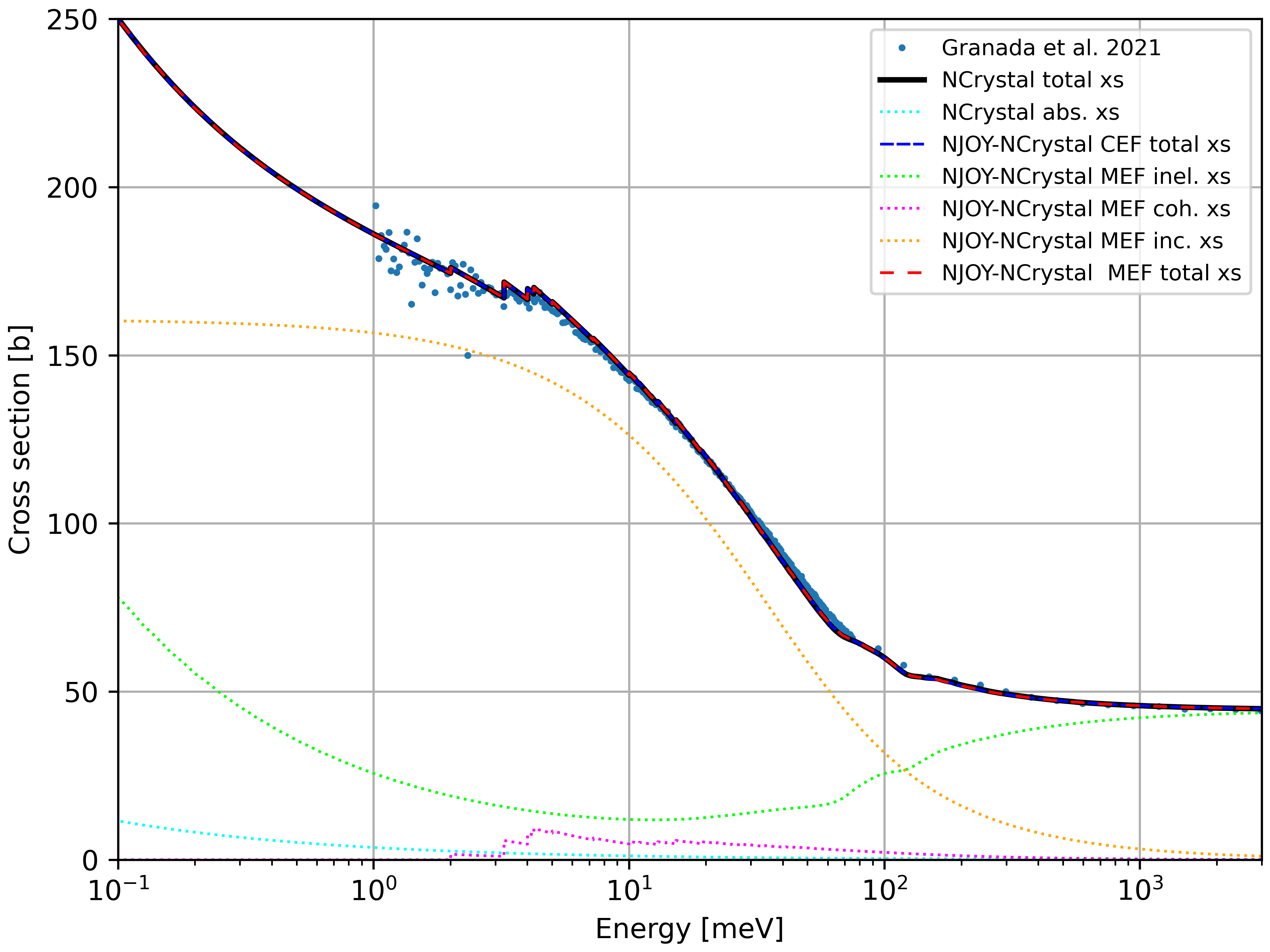}
  \caption{Total scattering cross section comparison for MgH$_2$.}
  \label{fig:mgh2_total_xs}
\end{figure}

From the calculated phonon spectra, tsl-ENDF files were created for both MgH$_2$ and MgD$_2$, using the NJOY~+~NCrystal tool. In Figure \ref{fig:mgh2_total_xs} it can be seen that the agreement between the measured total cross section data from Granada \cite{granada2021development} and the calculated curves (which include an absorption cross section ($\sigma_{2200}^\text{H} = 0.3326$ b, $\sigma_{2200}^\text{Mg} = 0.063$ b) is excellent. Since MgH$_2$ is an incoherent scatterer, due to the large incoherent cross section of hydrogen, \texttt{iel=99} is a good option as well. 

\begin{figure}[h!] 
  \centering
  \includegraphics[scale=0.5]{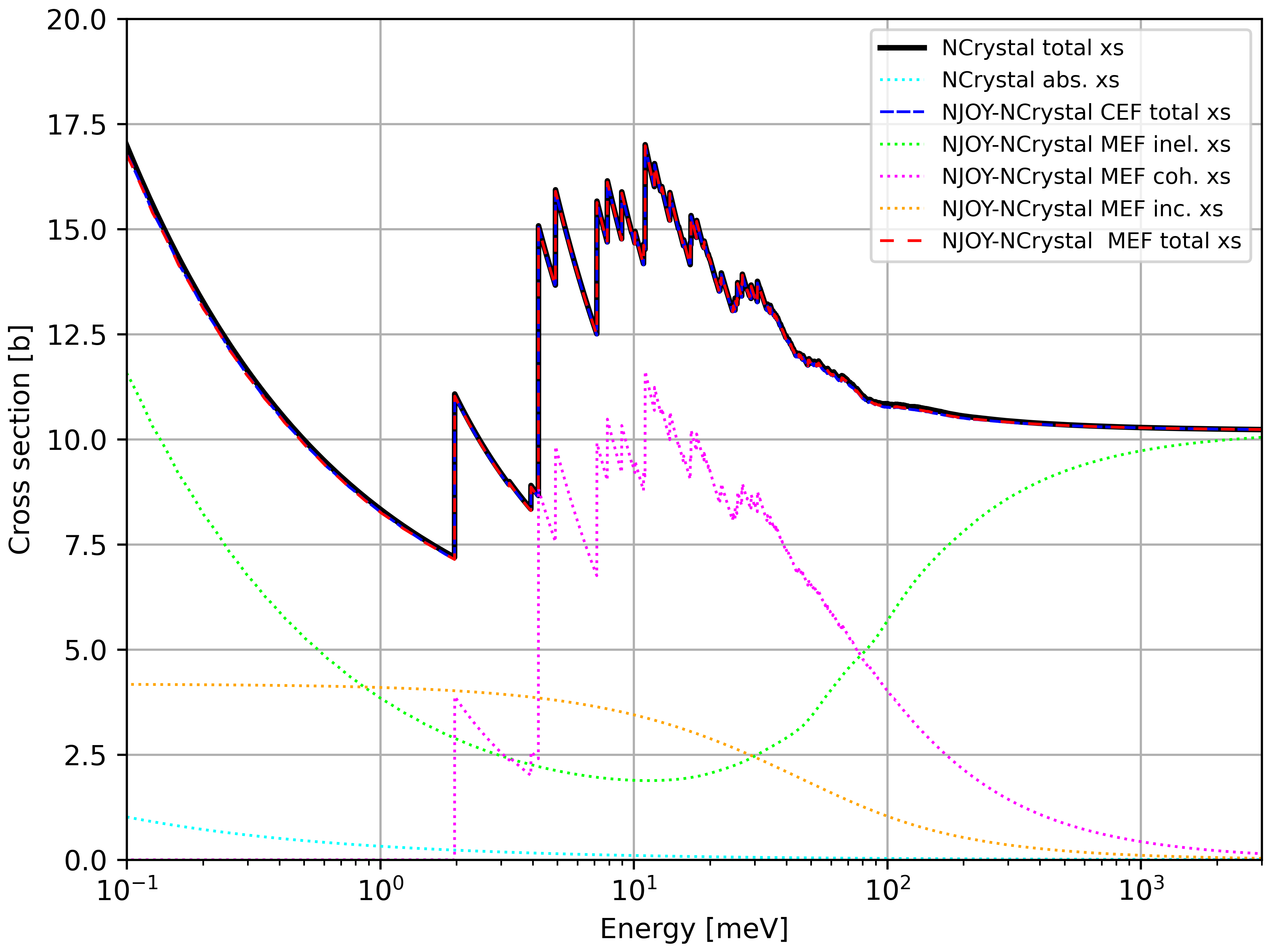}
  \caption{Total scattering cross section comparison for MgD$_2$.}
  \label{fig:mgd2_total_xs}
\end{figure}

Figure \ref{fig:mgd2_total_xs} shows a comparison between the NCrystal and the NJOY+NCrystal calculated total cross sections for MgD$_2$, as well as different scattering components as calculated in NJOY+NCrystal and NCrystal. The total cross section for MgD$_2$ is orders of magnitude lower than for MgH$_2$. This is due to incoherent hydrogen, with a cross section of 82.02 barns, being replaced with coherent deuterium with the cross section of 7.64 barns. From Figure \ref{fig:mgd2_total_xs} we can see that both the coherent and incoherent elastic components are not negligible in the thermal region and that the agreement between NCrystal, CEF and MEF calculated total cross sections is excellent.

\section{Conclusions}
In this work we presented a new tool for creating thermal neutron scattering libraries for crystalline solid materials that combines the strengths of the industry standard NJOY2016 with the new capabilities of the software library NCrystal. This new tool includes the implementation of the new mixed elastic format for ENDF-6 files, as well as a proposal for mixed elastic support in the .ACE format. The tool was used to create the largest collection of thermal scattering libraries to date. Explanations of changes made to NJOY2016, to create the tsl-ENDF and .ACE files, are presented as well as the changes to OpenMC to handle .ACE files in the new format. Examples for different types of scatterers have been provided (i.e. coherent-dominant, incoherent-dominant, as well as mixed elastic scatterers) coupled with experimental data validation. Phonon spectra have also been calculated and new tsl-ENDF files, for reflector materials being considered in the HighNESS project at the ESS, have been generated. 

Although NJOY+NCrystal represents significant improvement upon currently available methods for treatment of thermal neutron scattering in Monte Carlo particle codes, it is still limited by what can be stored in the ENDF-6 format. For a complete utilization of the capabilities of NCrystal, such as modelling single crystals, small angle neutron scattering, coherent one-phonon approximation, multi-phase and defects support, direct calls to NCrystal by Monte Carlo codes would need to be made. This will be the focus of future work.

All the source code presented in this work, as well as the thermal scattering libraries in tsl-ENDF and .ACE format are available in the Github repository of the HighNESS project: \url{https://github.com/highness-eu}.


\section{Supplemental information}
Supplemental information is available in the online version of this article, including plots of the total cross and heat capacity for all materials, and a text file containing the comments for each model. Numbers in the cross section plots are EXFOR entries. Experimental cross section data in the plots has not been renormalized.

\section{Acknowledgements}
This work was funded by the HighNESS project at the European Spallation Source. HighNESS is funded by the European Framework for Research and Innovation Horizon 2020, under grant agreement 951782.
\end{spacing}

\clearpage

\begin{table}[p]
  \caption{List of tsl-ENDF evaluations in the NJOY+NCrystal library.}
  \centering
  \label{table:endf_libraries}
  \begin{threeparttable}
  \begin{tabular}{p{0.125\linewidth}p{0.175\linewidth}>{\centering\arraybackslash}p{0.075\linewidth}p{0.125\linewidth}p{0.4\linewidth}}
  \toprule
  Material & Name & Material Number & Components & Temperature range, VDOS reference, Validation \\
    \midrule
AgBr & Silver Bromide & 1 & Br(AgBr), Ag(AgBr) & 20 - 700 K, \cite{wwwphonodb}, CS, SHC \\
 Ag & Silver & 2 & Ag & 20 - 1200 K, \cite{HART2017325}, CS, SHC, XS \\
 Al2O3 & Corundum & 3 & O(Al2O3), Al(Al2O3) & 20 - 2200 K, \cite{PhysRevB.67.174106}, CS, SHC, XS\\
 Al4C3 & Aluminium Carbide & 4 & C(Al4C3), Al(Al4C3) & 20 - 2400 K, \cite{wwwphonodb}, CS, SHC \\
 AlN & Aluminum Nitride & 5 & N(AlN), Al(AlN) & 20 - 2700 K, \cite{wwwphonodb}, CS, SHC \\
 Al & Aluminum & 6 & Al & 20 - 900 K, \cite{TOGO20151}, CS, SHC, XS \\
 Au & Gold & 7 & Au & 20 - 1200 K, \cite{LandoltBornstein1981:sm_lbs_978-3-540-38957-6_5}, CS, SHC, XS \\
 BaF2 & Barium Fluoride & 8 & F(BaF2), Ba(BaF2) & 20 - 1600 K, \cite{wwwphonodb}, CS, SHC \\
 BaO & Barium Oxide & 9 & O(BaO), Ba(BaO) & 20 - 2000 K, \cite{wwwphonodb}, CS, SHC \\
 Ba & Barium & 10 & Ba & 20 - 1000 K, \cite{BariumVDOS}, CS, SHC, XS  \\
 Be3N2 & Beryllium Nitride & 12 & Be(Be3N2), N(Be3N2) & 20 - 2400 K, \cite{wwwphonodb}, CS, SHC \\
 Bi2O3-$\beta$ & Bismuth Trioxide & 18 & O(Bi2O3-$\beta$), Bi(Bi2O3-$\beta$) & 600 - 900 K, \cite{wwwphonodb}, CS, SHC \\
 Bi & Bismuth & 22 & Bi & 20 - 500 K, \cite{KeinertBismuth}, CS, SHC, XS \\
 C-diamond & Diamond & 23 & C(C-diamond) & 20 - 4000 K, \cite{PhysRevB.60.9444}, CS, SHC  \\
 CF4-$\alpha$ & Carbon Tetrafluoride & 25 & C(CF4-$\alpha$), F(CF4-$\alpha$) & 20 - 70 K, \cite{wwwphonodb}, CS, SHC \\
 CaCO3 & Aragonite & 26 & C(CaCO3), O(CaCO3), Ca(CaCO3) & 20 - 1000 K, \cite{wwwphonodb}, CS, SHC \\
 CaF2 & Calcium Flouride & 27 & F(CaF2), Ca(CaF2) & 20 - 1600 K, \cite{wwwphonodb}, CS, SHC \\
 CaH2 & Calcium Hydride & 28 & H(CaH2), Ca(CaH2) & 20 - 1200 K, \cite{wwwphonodb}, CS, SHC \\
 CaOH2 & Calcium Hydroxide & 29 & H(CaOH2), O(CaOH2), Ca(CaOH2) & 20 - 800 K, \cite{wwwphonodb}, CS, SHC \\
 CaO & Calcium Oxide & 30 & O(CaO), Ca(CaO) & 20 - 2500 K, \cite{wwwphonodb}, CS, SHC \\
 CaZrO3 & Calcium Zirconate & 31 & O(CaZrO3), Ca(CaZrO3), Zr(CaZrO3) & 20 - 2500 K, \cite{wwwphonodb}, CS, SHC \\
 Ca & Calcium & 32 & Ca & 20 - 700 K, \cite{PhysRevB.27.3303},  CS, SHC, XS \\
 CeO2 & Cerium Oxide & 33 & O(CeO2), Ce(CeO2) & 20 - 2500 K, \cite{wwwphonodb}, CS, SHC \\
 Cr & Chromium & 34 & Cr & 20 - 2000 K, \cite{LandoltBornstein1981:sm_lbs_978-3-540-38957-6_13}, CS, SHC, XS  \\
 Cu2O & Cuprite & 35 & O(Cu2O), Cu(Cu2O) & 20 - 1400 K, \cite{wwwphonodb}, CS, SHC \\
 Cu & Copper & 36 & Cu & 20 - 1200 K. \cite{PhysRev.164.922}, CS, SHC, XS  \\
 Dy2O3 & Dysprosium Oxide & 37 & O(Dy2O3), Dy(Dy2O3) & 20 - 2400 K, \cite{wwwphonodb}, CS, SHC \\
 \bottomrule
 \end{tabular}
 \begin{tablenotes}\footnotesize
 \item[*] CS = Crystal structure
 \item[*] SHC = Specific heat capacity
 \item[*] XS = Total cross section measurement
 \end{tablenotes}
 \end{threeparttable}

\end{table}

\newpage

\begin{table*}[p]
      \caption{List of tsl-ENDF evaluations in the NJOY+NCrystal library (continued)}
\label{table:endf_libraries2}
  \centering
  \begin{threeparttable}
  \begin{tabular}{p{0.125\linewidth}p{0.175\linewidth}>{\centering\arraybackslash}p{0.075\linewidth}p{0.125\linewidth}p{0.4\linewidth}}
  \toprule
    Material & Name & Material Number & Components & Temperature range, VDOS reference, Validation \\
        \midrule
 Fe-$\alpha$ & Alpha Iron & 38 & Fe(Fe-$\alpha$) & 20 - 1100 K, \cite{PhysRev.162.528}, CS, SHC, XS\\
 Fe-$\gamma$ & Gamma Iron & 39 & Fe(Fe-$\gamma$) & 1200 - 1800 K, \cite{PhysRevB.35.4500}, CS, SHC \\
 Ge3Bi4O12 & Bismuth Germanate & 43 & O(Ge3Bi4O12), Ge(Ge3Bi4O12), Bi(Ge3Bi4O12) & 20 - 1300 K, \cite{wwwphonodb}, CS, SHC \\
 GeTe & Germanium Telluride & 44 & Ge(GeTe), Te(GeTe) & 20 - 600 K, \cite{wwwphonodb}, CS, SHC \\
 Ge & Germanium & 45 & Ge & 20 - 1200 K, \cite{PhysRevB.5.3151}, CS, SHC, XS \\
 Ho2O3 & Holmium Oxide & 47 & O(Ho2O3), Ho(Ho2O3) & 20 - 2400 K, \cite{wwwphonodb}, CS, SHC \\
 KBr & Potassium Bromide & 48 & K(KBr), Br(KBr) & 20 - 1000 K, \cite{wwwphonodb}, CS, SHC \\
 KF & Potassium Flouride & 49 & F(KF), K(KF) & 20 - 1100 K, \cite{wwwphonodb}, CS, SHC \\
 KOH & Potassium Hydroxide & 50 & H(KOH), O(KOH), K(KOH) & 20 - 500 K, \cite{wwwphonodb}, CS, SHC \\
 K & Potassium & 51 & K & 20 - 300 K, \cite{LandoltBornstein1981:sm_lbs_978-3-540-38957-6_24}, CS, SHC, XS \\
 La2O3 & Lanthanum Oxide & 52 & O(La2O3), La(La2O3) & 20 - 2500 K, \cite{wwwphonodb}, CS, SHC \\
 LaBr3 & Lanthanum Bromide & 53 & Br(LaBr3), La(LaBr3) & 20 - 1000 K, \cite{wwwphonodb}, CS, SHC \\
 Li2O & Lithium Oxide & 54 & Li(Li2O), O(Li2O) & 20 - 1700 K, \cite{wwwphonodb}, CS, SHC \\
 Li3N & Lithium Nitride & 55 & Li(Li3N), N(Li3N) & 20 - 1000 K, \cite{wwwphonodb}, CS, SHC \\
 LiF & Lithium Flouride & 56 & Li(LiF), F(LiF) & 20 - 1100 K, \cite{wwwphonodb}, CS, SHC \\
 LiH & Lithium Hydride & 57 & H(LiH), Li(LiH) & 20 - 900 K, \cite{wwwphonodb}, CS, SHC \\
 Lu2O3 & Lutetium Oxide & 58 & O(Lu2O3), Lu(Lu2O3) & 20 - 2400 K, \cite{wwwphonodb}, CS, SHC \\
 Mg2SiO4 & Magnesium Silicate & 59 & O(Mg2SiO4), Mg(Mg2SiO4), Si(Mg2SiO4) & 20 - 2100 K, \cite{wwwphonodb}, CS, SHC \\
 MgAl2O4 & Magnesium Aluminate Spinel & 60 & O(MgAl2O4), Mg(MgAl2O4), Al(MgAl2O4) & 20 - 2400 K, \cite{wwwphonodb}, CS, SHC \\
 MgCO3 & Magnesium Carbonate & 61 & C(MgCO3), O(MgCO3), Mg(MgCO3) & 20 - 600 K, \cite{wwwphonodb}, CS, SHC \\
 MgD2 & Magnesium Deuteride & 62 & D(MgD2), Mg(MgD2) & 20 - 600 K, CS, SHC, XS \\
 MgH2 & Magnesium Hydride & 64 & H(MgH2), Mg(MgH2) & 20 - 600 K \\
 MgOH2 & Magnesium Hydroxide & 65 & H(MgOH2), O(MgOH2), Mg(MgOH2) & 20 - 600 K, \cite{wwwphonodb}, CS, SHC \\
 Mg & Magnesium & 67 & Mg & 20 - 900 K, \cite{LandoltBornstein1981:sm_lbs_978-3-540-38957-6_26}, CS, SHC, XS  \\

 \bottomrule
    \end{tabular}
 \begin{tablenotes}\footnotesize
 \item[*] CS = Crystal structure
 \item[*] SHC = Specific heat capacity
 \item[*] XS = Total cross section measurement
 \end{tablenotes}
 \end{threeparttable}

\end{table*}

\newpage

\begin{table*}[p]
    \caption{List of tsl-ENDF evaluations in the NJOY+NCrystal library (continued)}
  \label{table:endf_libraries3}
  \centering
  \begin{threeparttable}
 \begin{tabular}{p{0.125\linewidth}p{0.175\linewidth}>{\centering\arraybackslash}p{0.075\linewidth}p{0.125\linewidth}p{0.4\linewidth}}
 \toprule
    Material & Name & Material Number & Components & Temperature range, VDOS reference, Validation \\
    \midrule
 Mo & Molybdenum & 68 & Mo & 20 - 2500 K, \cite{LandoltBornstein1981:sm_lbs_978-3-540-38957-6_27}, CS, SHC, XS \\
 Na$_4$Si$_3$Al$_3$O$_{12}$Cl & Sodalite & 69 & O(Sodalite), Na(Sodalite), Al(Sodalite), Si(Sodalite), Cl(Sodalite) & 20 - 1300 K, \cite{wwwphonodb}, CS, SHC \\
 NaBr & Sodium Bromide & 72 & Na(NaBr), Br(NaBr) & 20 - 2100 K, \cite{wwwphonodb}, CS, SHC \\
 NaCl & Sodium Chloride & 73 & Na(NaCl), Cl(NaCl) & 20 - 1000 K, \cite{wwwphonodb}, CS, SHC \\  
 NaF & Sodium Flouride & 74 & F(NaF), Na(NaF) & 20 - 1200, \cite{wwwphonodb}, CS, SHC K \\
 NaI & Sodium Iodide & 75 & Na(NaI), I(NaI) & 20 - 900 K, \cite{wwwphonodb}, CS, SHC \\
 NaMgH3 & Sodium Magnesium Hydride & 76 & H(NaMgH3), Na(NaMgH3), Mg(NaMgH3) & 20 - 600 K, \cite{wwwphonodb}, CS, SHC \\
 NaOH & Sodium Hydroxide & 77 & H(NaOH), O(NaOH), Na(NaOH) & 20 - 590 K, \cite{wwwphonodb}, CS, SHC \\
 Na & Sodium & 78 & Na & 20 - 350 K, \cite{LandoltBornstein1981:sm_lbs_978-3-540-38957-6_28}, CS, SHC, XS \\
 Nb & Niobium & 79 & Nb & 20 - 2500 K, \cite{LandoltBornstein1981:sm_lbs_978-3-540-38957-6_29}, CS, SHC, XS \\
 Nd2O3 & Neodymium Oxide & 80 & O(Nd2O3), Nd(Nd2O3) & 20 - 2400 K, \cite{wwwphonodb}, CS, SHC \\
 Ni & Nickel & 81 & Ni & 20 - 1500 K, \cite{PhysRevB.75.104301}, CS, SHC, XS \\
 P2O5 & Phosphorus Pentoxide & 82 & O(P2O5), P(P2O5) & 20 - 600 K, \cite{wwwphonodb}, CS, SHC \\
 Pb3O4 & Lead Oxide & 83 & O(Pb3O4), Pb(Pb3O4) & 20 - 800 K, \cite{wwwphonodb}, CS, SHC \\
 PbCO3 & Lead Carbonate & 84 & C(PbCO3), O(PbCO3), Pb(PbCO3) & 20 - 580 K, \cite{wwwphonodb}, CS, SHC \\
 PbF2 & Lead Flouride & 85 & F(PbF2), Pb(PbF2) & 20 - 1000 K, \cite{wwwphonodb}, CS, SHC \\
 PbO-$\alpha$ & Alpha Lead Monoxide & 87 & O(PbO-$\alpha$), Pb(PbO-$\alpha$) & 20 - 1100 K, \cite{wwwphonodb}, CS, SHC \\
 PbO-$\beta$ & Beta Lead Monoxide & 88 & O(PbO-$\beta$), Pb(PbO-$\beta$) & 20 - 1100 K, \cite{wwwphonodb}, CS, SHC \\
 PbS & Lead Sulfide & 89 & S(PbS), Pb(PbS) & 20 - 1300 K, \cite{wwwphonodb}, CS, SHC \\
 Pb & Lead & 90 & Pb & 20 - 600 K, \cite{GILAT1965101}, CS, SHC, XS \\
 Pd & Palladium & 91 & Pd & 20 - 1800 K, \cite{LandoltBornstein1981:sm_lbs_978-3-540-38957-6_32}, CS, SHC, XS \\
 Pt & Platinum & 93 & Pt & 20 - 2000 K, \cite{LandoltBornstein1981:sm_lbs_978-3-540-38957-6_33}, CS, SHC, XS \\
 Rb & Rubidium & 94 & Rb & 20 - 300 K, \cite{LandoltBornstein1981:sm_lbs_978-3-540-38957-6_34}, CS, SHC, XS \\
 Sc & Scandium & 96 & Sc & 20 - 1800 K, \cite{LandoltBornstein1981:sm_lbs_978-3-540-38957-6_38}, CS, SHC, XS \\
 SiC-$\alpha$ & Alpha Silicon Carbide & 98 & C(SiC-$\alpha$), Si(SiC-$\alpha$) & 1973 - 3100 K, \cite{wwwphonodb}, CS, DIFF \\
 SiC-$\beta$ & Beta Silicon Carbide & 99 & C(SiC-$\beta$), Si(SiC-$\beta$) & 20 - 1900 K \cite{wwwphonodb}, CS, DIFF \\
 \bottomrule
    \end{tabular}
\begin{tablenotes}\footnotesize
 \item[*] CS = Crystal structure
 \item[*] SHC = Specific heat capacity
 \item[*] XS = Total cross section measurement
 \item[*] DIFF = Diffraction data
 \end{tablenotes}
 \end{threeparttable}

\end{table*}

\newpage

\begin{table*}[p]
   \caption{List of tsl-ENDF evaluations in the NJOY+NCrystal library (continued)}
  \label{table:endf_libraries4}
 \centering
 \begin{threeparttable}
 \begin{tabular}{p{0.125\linewidth}p{0.175\linewidth}>{\centering\arraybackslash}p{0.075\linewidth}p{0.125\linewidth}p{0.4\linewidth}}
 \toprule
    Material & Name & Material Number & Components & Temperature range, VDOS reference, Validation \\
    \midrule
 SiLu2O5 & Silicon Lutetium Oxide & 100 & O(SiLu2O5), Si(SiLu2O5), Lu(SiLu2O5) & 20 - 2400 K, \cite{wwwphonodb}, CS, SHC \\
SiO2-$\alpha$ & Alpha Quartz & 101 & O(SiO2-$\alpha$), Si(SiO2-$\alpha$) & 20 - 800 K, \cite{wwwphonodb}, CS, SHC, XS \\
 SiO2-$\beta$ & Beta Quartz & 102 & O(SiO2-$\beta$), Si(SiO2-$\beta$) & 846.15 - 1900 K, \cite{wwwphonodb}, CS, SHC \\
 SiY2O5 & Silicon Yttrium Oxide & 103 & O(SiY2O5), Si(SiY2O5), Y(SiY2O5) & 20 - 2300 K, \cite{wwwphonodb}, CS, SHC \\
 Si & Silicon & 104 & Si & 20 - 1600 K, \cite{Liu2014}, CS, SHC, XS \\
 Sn & Tin & 107 & Sn & 20 - 500 K, \cite{PhysRevB.61.R14881}, CS, SHC, XS \\
 SrF2 & Strontium Fluoride & 108 & F(SrF2), Sr(SrF2) & 20 - 1400 K, \cite{wwwphonodb}, CS, SHC \\
 SrH2 & Strontium Hydride & 109 & H(SrH2), Sr(SrH2) & 20 - 1300 K, \cite{wwwphonodb}, CS, SHC \\
 Sr & Strontium & 111 & Sr & 20 - 1000 K, \cite{PhysRevB.30.3502}, CS, SHC \\
 Th3N4 & Thorium Nitride & 112 & N(Th3N4), Th(Th3N4) & 20 - 3000 K, \cite{wwwphonodb}, CS, SHC \\
 ThO2 & Thorium Dioxide & 113 & O(ThO2), Th(ThO2) & 20 - 3500 K, \cite{wwwphonodb}, CS, SHC \\
 ThSiO4 & Huttonite & 114 & O(ThSiO4), Si(ThSiO4), Th(ThSiO4) & 20 - 2000 K, \cite{wwwphonodb}, CS, SHC \\
 TiO2-anatase & Anatase & 115 & O(TiO2-anatase), Ti(TiO2-anatase) & 20 - 2000 K, \cite{wwwphonodb}, CS, SHC \\
 TiO2-rutile & Rutile & 116 & O(TiO2-rutile), Ti(TiO2-rutile) & 873 - 2000 K, \cite{wwwphonodb}, CS, SHC \\
 Ti & Titanium & 117 & Ti & 20 - 1800 K, \cite{PhysRevB.19.181}, CS, SHC, XS  \\
 TlBr & Thalium Bromide & 118 & Br(TlBr), Tl(TlBr) & 20 - 700 K, \cite{wwwphonodb}, CS, SHC \\
 Tm2O3 & Thulium Oxide & 119 & O(Tm2O3), Tm(Tm2O3) & 20 - 2400 K, \cite{wwwphonodb}, CS, SHC \\
 UF6 & Uranium Hexaflouride & 120 & F(UF6), U(UF6) & 20 - 320 K, \cite{wwwphonodb}, CS, SHC \\
 V & Vanadium & 122 & V & 20 - 2000 K, \cite{PhysRevLett.107.115501}, CS, SHC, XS  \\
 W & Tungsten & 123 & W & 20 - 3500 K, \cite{Crocombette_2015}, CS, SHC, XS  \\

\bottomrule

\end{tabular}
\begin{tablenotes}\footnotesize
 \item[*] CS = Crystal structure
 \item[*] SHC = Specific heat capacity
 \item[*] XS = Total cross section measurement
 \end{tablenotes}
 \end{threeparttable}
\end{table*}

\newpage

\begin{table}[p]
   \caption{List of tsl-ENDF evaluations in the NJOY+NCrystal library (continued)}
  \label{table:endf_libraries5}
 \centering
 \begin{threeparttable}
 \begin{tabular}{p{0.125\linewidth}p{0.175\linewidth}>{\centering\arraybackslash}p{0.075\linewidth}p{0.125\linewidth}p{0.4\linewidth}}
    Material & Name & Material Number & Components & Temperature range, VDOS reference, Validation \\
    \midrule
 Y2O3 & Yttrium Oxide & 124 & O(Y2O3), Y(Y2O3) & 20 - 2500 K, \cite{PhysRevB.84.094301}, CS, SHC \\
 Y3Al5O12 & Yttrium Aluminium Garnet & 125 & O(Y3Al5O12), Al(Y3Al5O12), Y(Y3Al5O12) & 20 - 2900 K, \cite{wwwphonodb}, CS, SHC \\
 YAlO3 & Yttrium Orthoaluminate & 126 & O(YAlO3), Al(YAlO3), Y(YAlO3) & 20 - 2000 K, \cite{wwwphonodb}, CS, SHC \\
 Y & Yttrium & 127 & Y & 20 - 1600 K, \cite{LandoltBornstein1981:sm_lbs_978-3-540-38957-6_50}, CS, SHC, XS  \\
 ZnF2 & Zinc Flouride & 128 & F(ZnF2), Zn(ZnF2) & 20 - 1000 K, \cite{wwwphonodb}, CS, SHC \\
 ZnO & Zinc Oxide & 129 & O(ZnO), Zn(ZnO) & 20 - 2200 K, \cite{wwwphonodb}, CS, SHC \\
 ZnS-sphalerite & Zinc Sulfide & 130 & S(ZnS-sphalerite), Zn(ZnS-sphalerite) & 20 - 1300 K, \cite{wwwphonodb}, CS, SHC \\
 Zn & Zinc & 131 & Zn & 20 - 600 K, \cite{LandoltBornstein1981:sm_lbs_978-3-540-38957-6_51}, CS, SHC, XS  \\
 ZrF4-$\beta$ & Beta Zirconium Tetrafluoride & 132 & F(ZrF4-$\beta$), Zr(ZrF4-$\beta$) & 773 - 1150 K, \cite{wwwphonodb}, CS, SHC \\
 ZrO2-tet & Tetragonal Zirconium Dioxide & 133 & O(ZrO2-tet), Zr(ZrO2-tet) & 1443.15 - 2600 K, \cite{wwwphonodb}, CS, SHC \\
 ZrO2 & Zirconium Dioxide & 134 & O(ZrO2), Zr(ZrO2) & 20 - 1400 K, \cite{wwwphonodb}, CS, SHC \\
 ZrSiO4 & Zirconium Orthosilicate & 135 & O(ZrSiO4), Si(ZrSiO4), Zr(ZrSiO4) & 20 - 1800 K, \cite{wwwphonodb}, CS, SHC \\
 Zr & Zirconium & 136 & Zr & 20 - 2000 K, \cite{Crocombette_2015}, CS, SHC, XS  \\

\bottomrule

\end{tabular}
\begin{tablenotes}\footnotesize
 \item[*] CS = Crystal structure
 \item[*] SHC = Specific heat capacity
 \item[*] XS = Total cross section measurement
 \end{tablenotes}
 \end{threeparttable}

\end{table}

\clearpage

\begin{spacing}{2.0}
\bibliographystyle{unsrtnat}
\bibliography{cas-refs}
\end{spacing}



\end{document}